\documentclass[%
 reprint,
nofootinbib,
 amsmath,amssymb,
 aps,
]{revtex4-1}

\usepackage{graphicx}
\usepackage{dcolumn}
\usepackage{bm}
\usepackage{hyperref}

\usepackage{braket}
\usepackage{mathtools}
\usepackage{enumitem}
\usepackage{soul} 
\usepackage{tikz}
\usetikzlibrary{shapes,arrows}
\usepackage{comment} 

\newcommand\FracLineWidth{0.9}

\begin{document}

\preprint{APS/123-QED}

\title{
Quantum state preparation of spin eigenstates including the Dicke states with generalized all-coupled interaction in a spintronic quantum computing architecture\\
}

\author{Amritesh Sharma}
 \email{amritesh.iitb@gmail.com}
\author{Ashwin A. Tulapurkar}%
 \email{ashwin@ee.iitb.ac.in}
\affiliation{%
Solid State Devices Group, Department of Electrical Engineering, Indian Institute of Technology - Bombay, Mumbai 400076, India
}%


\begin{abstract}


There has been an extensive development in the use of multi-partite entanglement as a resource for various quantum information processing tasks. In this paper we focus on preparing arbitrary spin eigenstates whose subset contain important entangled resources like Dicke states as well as some other sub-radiant states that are  difficult to prepare.
Leveraging on the symmetry of these states we consider uniform pairwise exchange coupling between every pair of qubits. Starting from a product state of a given spin eigenstate with a single qubit state, another spin eigenstate  can be prepared using simple time evolutions. This expansion paves a deterministic approach to prepare arbitrary Dicke states in linear steps. We discuss an improvement in this cost building up on a previous work for W states deterministic preparation in logarithmic circuit depth \cite{LogWstate}. The modified algorithm requires several iterations of pumping spin angular momentum into the system and is akin to the amplitude amplification in Grover’s search. 
As a use case to demonstrate the proposed scheme, we choose a system of non-interacting static spin qubits connected to a ferromagnetic reservoir. The flying qubits emerging from the reservoir locally interact with static qubits successively, mediating an in-direct exchange interaction between all the pairs. 







\end{abstract}

\pacs{Valid PACS appear here}
\maketitle


\section{\label{sec:intro} Introduction}

%
%

Efficiently preparing arbitrary quantum states is a challenging task corroborated by exponentially growing Hilbert space with the size of the system \citep{ShendeSynth2006}. Among the exponential variety, states with sufficient structure and symmetry are relatively easier to investigate and prepare \citep{LogWstate}. Those exhibiting properties like multi-partite entanglement are extremely useful and there has been an extensive development in their use as a resource for quantum information processing tasks.

In a seminal paper in 1954, R.H. Dicke introduced the idea of super-radiant states, commonly referred to as Dicke states in literature \citep{Dicke1954}. These states find potential applications in realization of small linewidth superradiant lasers \citep{Meiser_2009}, enhancement of spin-photon coupling in cavity QED systems \citep{Breeze_2017} for investigating many body systems \citep{Baumann_2010, Yu_2019} and so on. Their entanglement properties like robustness to particle losses \citep{Neven_2018} and immunity to collective dephasing noise \citep{Lidar_2003} have been studied making them useful for several quantum computing and communication applications \citep{Kiesel_2007, Sen_De__2003, Childs_cliques, Ivanov_2010, Prevedel_2009, Chiuri_2012, Toth:07, T_th_2012}. 
As such their preparation schemes have been investigated a lot, with some experimental demonstrations, in many promising physical systems of the NISQ era like trapped ion \citep{Ivanov_2013,Hume_2009}, cavity and circuit QED systems \citep{Ji_2019,Xiao_2007,Wu_2017}, photonics \citep{Wieczorek_2009,Prevedel_2009, Wang_2016} and silicon (based on Kane quantum computing architecture) \citep{Luo_2012}. 
Ref.~\citep{B_rtschi_2019} proposes a circuit for deterministic preparation that is also suitable for quantum compression.
Another category of states called sub-radiant states are also-well studied and their preparation is also a challenging task \citep{SubRadWallraff, SubRadDeVoe, SubRadLiber}. It is believed that these might have applications in quantum memories \citep{Begzjav_2019}. The spin eigenfunctions encompasses these states 
and developing a deterministic algorithm of their preparation is the focus of this paper.



We shall also consider a spintronic use case to demonstrate our algorithm that is suitable for quantum computation \citep{sutton2015manipulating, kulkarni2019spin} and has been studied previously with single and two-qubit universal gate sets. The essential idea is that interaction of flying spin-qubit with a chain of static qubits can lead to entanglement of the static qubits. Similarly, if the flying qubit interacts with a single static qubit, it can change the quantum state of the static qubit.

 
  This is similar to 'classically' manipulating the orientation of a nano-magnet\citep{Shuklaeabc2618}  using travelling spin polarized electrons \citep{slonczewski1996current, berger1996emission}. These electrons are usually provided via spin polarized currents and there exist ample ways to generate these currents like spin-pumping \citep{bhuktare2019direct}, spin Hall effect \citep{bose2017sensitive, bose2018observation}, spin-dependent thermoelectric effects \citep{bose2016observation}, spin Nernst effect \citep{bose2019recent, bose2018direct} to name a few and hence observe this phenomena in literature. This phenomena is called the spin transfer torque and the terminology has been innocuously carried over to the quantum scales as well. Our adaptation enables the direct realization of non-decomposed (into smaller sized single or two qubit gates) multi-qubit gates that can enable lower-circuit depth implementations for certain algorithms and is in spirit of architecture-awareness. 

The algorithm we propose here is composed of expansion steps, where a single qubit in 0 or 1 state is appended to an n-1 qubit spin eigenstate and unitarily evolved for appropriate time to yield another n-qubit spin eigenstate accurate upto relative phase factors between certain (chosen) basis states in the superposition. These are corrected via single-qubit operations. The states so obtained obey the rules of angular momentum addition and correspond to 
genealogically indexed spin-eigen states \citep{Pauncz_1979}. 
We suggest an improvement in the scope of these expansions schemes to facilitate sub-linear-depth circuits. 
The unitary evolution considered here relies on identical pair-wise exchange interaction between any pair of qubits and it is understood such all-to-all connectivity can quickly become a bottleneck scaling up the physical system in a direct implementation. We have shown in our previous work \citep{LogWstate} that it is possible to engineer such a coupling indirectly using ancillary qubits in a "one-spin-down" subspace. We find that similar design is possible in other subspaces as well but the design instead emulates a generalization of the all coupled Hamiltonian, explained in the main text, that also enables the algorithm we propose with enough accuracy to yield high fidelity states: we shall demonstrate the preparation of Dicke states in this scenario.

\section{\label{sec:PrepScheme} Proposed Preparation Scheme}

Let us start by discussing the deterministic method of spin-eigenstates states expansion. Spin-eigenstates are defined as simultaneous eigenfunctions of $\boldsymbol{S}^2$ and $\boldsymbol{S}_z$ operators. 

\subsection{Expansion Methods} {\label{subsec:ExpansionMethod}

The total spin angular momentum $\boldsymbol{S}$ is defined as the tensor sum, $\boldsymbol{S}=\sum_{i} \boldsymbol{S}_i$, where $\boldsymbol{S}_i$ denote spin operator of $i^{th}$ qubit. We will 
denote the spin eigenstates by $\ket{X(n,S,M)}$, where n is number of qubits, and  S and M denote the spin and z-component of spin, quantum numbers i.e. $\boldsymbol{S}^2  \ket{X(n,S,M)}=S(S+1)\hbar^2\ket{X(n,S,M)}$ and $\boldsymbol{S}_z  \ket{X(n,S,M)}=M\hbar \ket{X(n,S,M)}$. We will denote the two single qubit states, $S=1/2, M=\pm 1/2$ by $\ket{0}$ and $\ket{1}$ respectively.
The eigenstates of n-1 qubit system with spin S and a single qubit can be combined to yield eigenstates with spin of $S\pm (1/2)$.
This is expressed in Eq.~\ref{eq:Ruben1} (See Ref.~\citep{Pauncz_1979}) where we have defined $A=\sqrt{\frac{S+M+1}{2S+1}}$ and $B=\sqrt{\frac{S-M}{2S+1}}$. The state is normalized as $A^2+B^2=1$. It should be noted that the values of n,S and M may not specify a unique state, and this issue is dealt with later. The  equations Eq.~\ref{eq:Ruben1} a and b can be inverted as given in Eq.~\ref{eq:Ruben2}

\begin{widetext}
\begin{subequations} \label{eq:Ruben1}
\begin{eqnarray}
\ket{X(n,S+\frac{1}{2},M+\frac{1}{2})}=A\ket{X(n-1,S,M)}\otimes \ket{0} +B\ket{X(n-1,S,M+1)}\otimes\ket{1}\\
\ket{X(n,S-\frac{1}{2},M+\frac{1}{2})}=-B\ket{X(n-1,S,M)}\otimes \ket{0}+A\ket{X(n-1,S,M+1)}\otimes\ket{1}  
\end{eqnarray}
\end{subequations}

\begin{subequations} \label{eq:Ruben2}
\begin{eqnarray}
\ket{X(n-1,S,M)}\otimes\ket{0}=A\ket{X(n,S+\frac{1}{2},M+\frac{1}{2})}-B \ket{X(n,S-\frac{1}{2},M+\frac{1}{2})}
\\
\ket{X(n-1,S,M+1)}\otimes\ket{1}=B\ket{X(n,S+\frac{1}{2},M+\frac{1}{2})}+A\ket{X(n,S-\frac{1}{2},M+\frac{1}{2})}
\end{eqnarray}
\end{subequations}
\end{widetext}


We now assume that we have two kinds of Hamiltonians at our disposal and show that spin eigenfunction of n-qubit system can be prepared from n-1 qubit system by subjecting the later system to time evolutions under these two Hamiltonians. The two Hamiltonians we need are: 1) Heisenberg exchange interaction between all pairs of qubits 2) Zeeman interaction generated by local magnetic fields along z direction acting on the qubits. We have shown in Ref.~\citep{LogWstate} that both these Hamiltonians can be engineered in a system of non-interacting spin qubits connected to ferromagnetic reservoirs. This aspect is further discussed in section \ref{sec:sTT}. It should be noted that the first Hamiltonian can be used for entangling the qubits, while the second Hamiltonian can be used for single qubit z-axis rotation operation. The first Hamiltonian can be written as:
\begin{equation} \label{eq:Ideal_Hamiltonain1}
\mathcal{H} = J^{\prime} \sum_{i<j}  \bm{S_i} \cdot \bm{S_j}= \frac{J^{\prime}}{2}  (\bm{S}^2 - \sum^n_{i=1} \bm{S}_i^2)
\end{equation}
where $J^{\prime}$ denotes the exchange interaction strength. The above equation shows that eigenstates $\boldsymbol{S}^2$ operator are also eigenstates of  all coupled Heisenberg Hamiltonian with eigenvalue $(J^{\prime} \hbar^2/2)[S(S+1)-3n/4]$. 
Let us now see how we can prepare state  $\ket{X(n,S+1/2,M+1/2)}$ from $\ket{X(n-1,S,M)}$ state. We append a qubit in $\ket{0}$ state to the system and  take the initial state as  a tensor product state $\ket{\psi (0)}=\ket{X(n-1,S,M)}\otimes\ket{0}$. We now subject the initial state to time evolution under the all coupled Heisenberg interaction. As the states $\ket{X(n,S\pm 1/2,M+1/2}$ are eigenfunctions of the Hamiltonian, the state at time $t$ can be written as, 
\begin{equation} \label{eq:c1_c2}
\begin{aligned}
\ket{\psi (t)}= c_1(t)\ket{X(n,S+1/2,M+1/2}\\+c_2(t) \ket{X(n,S-1/2,M+1/2)}
\end{aligned}
\end{equation}
with $c_1(0)=A$ and $c_2(0)=-B$. From equation \ref{eq:Ruben1}a and \ref{eq:Ruben1}b, we see that both the spin-eigenstates are linear combinations of $\ket{X(n-1,S,M)} \otimes \ket{0}$ and $\ket{X(n-1,S,M+1)} \otimes \ket{1}$. We can therefore write the state at time $t$ as: 
\begin{equation} \label{eq:a1_a2}
\begin{aligned}
\ket{\psi (t)}=a_1(t) \ket{X(n-1,S,M)} \otimes \ket{0} \\+a_2(t)\ket{X(n-1,S,M+1)}\otimes\ket{1}]
\end{aligned}
\end{equation}
with appropriately normalized $a_1$ and $a_2$ factors. We can easily get the expressions for $a_1$ and $a_2$ in terms of $c_1$ and $c_2$ (and vice-versa) from equations \ref{eq:Ruben1} and \ref{eq:Ruben2}, as given below:
\begin{equation} \label{eq:Ruben3}
\begin{aligned}
a_1=A c_1-B c_2, a_2=B c_1+A c_2 \\
c_1=A a_1+B a_2 , c_2=-B a_1+A a_2 
\end{aligned}
\end{equation}

Under the evolution with all coupled Heisenberg interaction, ignoring the global phase, we can write $c_1(t)=\exp(-i\omega t)c_1(0)$ and $c_2(t)=c_2(0)$, where $\omega=(E_1-E_2)/\hbar$ and $E_1$ and $E_2$ are eigenvalues of $\ket{X(n,S+1/2,M+1/2)}$ and $\ket{X(n,S-1/2,M+1/2)}$ states respectively. $\omega$ is given by, $\omega=J^{\prime} (S+1/2) \hbar$. We can obtain $a_1(t)$ and $a_2(t)$ from equation \ref{eq:Ruben3} as $a_1(t)=A^2[\exp(-i\omega t)-1]+1$ and $a_2(t)=A B[\exp(-i\omega t)-1]$. We stop the time evolution at time $t_s$, when the amplitudes are given by  $|a_1(t_s)|=A$ and $|a_2(t_s)|=B$. ( Normalization of the state imply that if $|a_1|=A$ then $|a_2|=B$ and vice-versa). Using the above expressions for $a_1$ or $a_2$, we get the following equation for $t_s$:
\begin{equation} \label{eq:ts_case1}
\cos(\omega t_s)=1-\frac{1}{2A^2}=1-\frac{1}{2}\frac{2S+1}{S+M+1}
\end{equation}
The state time time $t_s$ is very close to the desired $\ket{X(n,S+1/2,M+1/2)}$ state except for a relative phase factor. The state $\ket{\psi(t_s)}$, ignoring the global phase,  can be written as:
\begin{equation} \label{eq:Ruben5}
\begin{aligned}
\ket{\psi(t_s)}=A \ket{X(n-1,S,M)}\otimes \ket{0} \\+e^{i\phi}B\ket{X(n-1,S,M+1)}\otimes\ket{1}
\end{aligned}
\end{equation}
The relative phase factor $e^{i\phi} = (A/B)(a_2(t_s)/a_1(t_s))$ can be easily corrected by applying local magnetic field to the last qubit for a certain time. This corresponds to application of $R_z(\theta)$ gate on the last qubit, which performs a rotation about z-axis, modifies the relative phase factor to $e^{i(\phi+\theta)}$ in Eq.~\ref{eq:Ruben5}. This clearly leads to the desired spin eigenstate for a rotation amount $\theta = 2m\pi - \phi$ for any integer m.
Eq.~\ref{eq:ts_case1} besides providing the value of $t_s$ also gives the condition on S and M for which the algorithm would work. It can be easily checked that for real values $t_s$, the equation can be satisfied only when $2S+4M+3\geq 0$. 
In the above, we prepared $\ket{X(n,S+1/2,M+1/2)}$ state staring from$\ket{X(n-1,S,M)}$ state and appending a qubit in $\ket{0}$ state. The final state essentially corresponds to the first term in Eq.~\ref{eq:Ruben2}a. By appropriately choosing the stopping time $t_s$, followed by single qubit rotation of the last qubit, we can as well prepare $\ket{X(n,S-1/2,M+1/2)}$ state which corresponds to the second term in Eq.~\ref{eq:Ruben2}a. The stopping time in this case is chosen by to satisfy the condition $|a_1(t_s)|=B$ or $|a_2(t_s)|=A$. This condition gives the following equation for $t_s$:
\begin{equation} \label{eq:ts_case2}
\cos(\omega t_s)=1-\frac{1}{2B^2}=1-\frac{1}{2}\frac{2S+1}{S-M}
\end{equation}
The state at time $t_s$ is very close to the desired $\ket{X(n,S-1/2,M+1/2)}$ state except for a phase factor. The state $\ket{\psi(t_s)}$ can be written as
\begin{equation} \label{eq:Ruben5_case2}
\begin{aligned}
\ket{\psi(t_s)}=-B \ket{X(n-1,S,M)}\otimes \ket{0} \\+e^{i\phi} A\ket{X(n-1,S,M+1)}\otimes\ket{1}
\end{aligned}
\end{equation}
As before, the relative phase factor can be corrected by single qubit rotation on the last qubit. The algorithm would work for real values of $t_s$, which gives the condition that $2(S-2M) \geq 1$.

Now if we examine Eq.~\ref{eq:Ruben2}b, we see that we can append a qubit in $\ket{1}$ state (instead of $\ket{0}$ state) to the n-1 qubit spin-eigenstate  and prepare n qubit spin-eigenstates. The algorithm is again same as before: The system is subjected to the evolution under all coupled Heisenberg Hamiltonian for a certain time followed by phase correction by single qubit rotation operation on the last qubit. We will write the same equations as \ref{eq:c1_c2} and \ref{eq:a1_a2} with difference that $c_1(0)=B$ and $c_2(0)=A$. Putting $c_1(t)=c_1(0) \exp(-i\omega t)$ and $c_2(t)=c_2(0)$ in equation \ref{eq:Ruben3}, we get $a_1(t)=A B[\exp(-i\omega t)-1]$ and $a_2(t)=B^2[\exp(-i\omega t)-1]+1$. We stop the time evolution at time $t_s$, when the amplitudes are given by  $|a_1(t_s)|=A$ and $|a_2(t_s)|=B$ if desired final state is $\ket{X(n,S+1/2,M+1/2)}$. We get:
\begin{equation} \label{eq:ts_case3}
\cos(\omega t_s)=1-\frac{1}{2B^2}=1-\frac{1}{2}\frac{2S+1}{S-M}
\end{equation}
If the final desired state is $\ket{X(n,S-1/2,M+1/2)}$, the time evolution is stopped when $|a_1(t_s)|=B$ and $|a_2(t_s)|=A$. We get:
\begin{equation} \label{eq:ts_case4}
\cos(\omega t_s)=1-\frac{1}{2A^2}=1-\frac{1}{2}\frac{2S+1}{S+M+1}
\end{equation}

Let us now examine the case where spin quantum number is increased by 1/2 i.e. preparation of $S+1/2$ spin state from $S$ spin-state. We shall call it the spin incrementing method from now on. This can be done in two ways as discussed above: We can prepare $\ket{X(n,S+1/2, M+1/2)}$ state from $\ket{X(n-1,S, M)}$ or $\ket{X(n-1,S, M+1)}$ by appending $\ket{0}$ or $\ket{1}$ respectively. Calling the final state quantum numbers as $\widetilde{S}=S+1/2$ and $\widetilde{M}=M+1/2$, and using  equations \ref{eq:ts_case1} and \ref{eq:ts_case3}, we see that the two schemes work under the respective conditions: $\widetilde{S}+2\widetilde{M}\geq 0$ and  $\widetilde{S}-2\widetilde{M}\geq 0$. It is thus always possible to reach the final state using one of these two or both expansion schemes. We can similarly examine the case where the final state spin quantum number is decreased by 1/2. In the same spirit, we can call this the spin decrementing method. This again can be done in two ways: We can prepare $\ket{X(n,S-1/2, M+1/2)}$ state from $\ket{X(n-1,S, M)}$ or $\ket{X(n-1,S, M+1)}$ by appending $\ket{0}$ or $\ket{1}$ respectively. Calling the final state quantum numbers as $\widetilde{S}=S-1/2$ and $\widetilde{M}=M+1/2$, and using  equations \ref{eq:ts_case2} and \ref{eq:ts_case4}, we see that the two algorithms work under the respective conditions  $\widetilde{S}-2\widetilde{M}+1\geq 0$ and $\widetilde{S}+2\widetilde{M}+1\geq 0$. It is thus always possible to reach the final state using one of these two or both algorithms.
 
We now address the problem mentioned earlier that n,S and M quantum numbers may not be sufficient to specify a state completely. We can use a "branching diagram" to represent a spin-eigenstate. (See Ref.~\cite{Pauncz_1979} for details). We essentially construct a state starting from one qubit state and increasing the number of electrons in each stage, using either equation \ref{eq:Ruben1} a or \ref{eq:Ruben1} b. Thus in each stage we either increase or decrease spin by 1/2. Specifying the path of how a state is constructed, specifies the state uniquely. Increasing spin step is denoted by 1 and decreasing spin step is denoted by 2. As an example, let's consider state $\ket{X(n=5,S=3/2,M=1/2,11211)}$. The state $\ket{X(n=5,S=3/2,M=1/2)}$ has a degeneracy of 5 and path 11211 is required to specify the state uniquely (This state turns out be $\sqrt{(1/18)}[2\ket{00101} +2 \ket{00110} -\ket{01001}-\ket{01010}+\ket{01100}-\ket{10001}-\ket{10010}+\ket{10100} -2 \ket{11000}]$). To prepare this state on a quantum computer (equipped with the two Hamiltonians mentioned before), we essentially follow the same path viz. 11211. In each step there could be two possible ways to prepare the next state. Let's trace back the path starting with the final state.  We need to follow path of increasing spin i.e. we can prepare the final state from $\ket{X(n=4,S=1,M=0,1121)}$ or $\ket{X(n=4,S=1,M=1,1121)}$ by appending $\ket{0}$ or $\ket{1}$ respectively. Both of these options are allowed as they satisfy $\widetilde{S}+2\widetilde{M} \geq 0$ and $\widetilde{S}-2\widetilde{M} \geq 0$ conditions. Let's choose to prepare the final state from $\ket{\psi_4}=\ket{X(n=4,S=1,M=0,1121)}$. To prepare $\ket{\psi_4}$ we need to follow the path of increasing spin, which means that it can be prepared from $\ket{X(n=3,S=1/2,M=-1/2,112)}$ or $\ket{X(n=3,S=1/2,M=1/2,112)}$ by appending $\ket{0}$ or $\ket{1}$ respectively. Both the options are allowed. Let's choose to prepare the $\ket{\psi_4}$ state from $\ket{\psi_3}=\ket{X(n=3,S=1/2,M=1/2,112)}$. To prepare $\ket{\psi_3}$, we need to follow spin decreasing path i.e. the state can be prepared from $\ket{X(n=2,S=1,M=0,11)}$ or $\ket{X(n=2,S=1,M=1,11)}$ by appending $\ket{0}$ or $\ket{1}$ respectively. Again both these options are allowed. If we choose $\ket{\psi_2}=\ket{X(n=2,S=1,M=1,11)}$, then we have finished the process as $\ket{\psi_2}=\ket{00}$ is a completely un-entangled state. If we choose $\ket{\psi_2}=\ket{X(n=2,S=1,M=0,11)}$, we need one more step to prepare $\ket{\psi_2}$ from $\ket{1}$ or $\ket{0}$ by following spin increasing path.
It is clear from above arguments that we can prepare a general spin-eigenstate of n-qubit system in O(n) stages with each stage involving an entangling evolution under all coupled Heisenberg interaction followed by single qubit rotation operation. We shall see later how this linear step algorithm can be improved for a special class of spin-eigenstates called Dicke states.
\subsection{Generalizing Expansion Methods: Amplitude ampification} \label{subsec:AmplitudeAmp}
Let's consider preparation of state $\ket{X(n,S=n/2,M)}$. This is the state with highest value of spin and the path to be followed to prepare it, is always spin-increasing i.e. we do not need to specify the path separately. As a particular example, let's choose $M=S-2=(n-4)/2$. This state can be prepared from $\ket{X(n-1,S=(n-1)/2,M=(n-5)/2}$ or $\ket{X(n-1,S=(n-1)/2,M=(n-3)/2}$ by appending $\ket{0}$ or $\ket{1}$ respectively. However, state $\ket{X(n-1,S=(n-1)/2,M=(n-3)/2}$ can not be used as it does not satisfy the condition $\widetilde{S}-2\widetilde{M}\geq 0$. For the same reason, even the state $\ket{X(n-1,S=(n-1)/2,M=(n-3)/2}$ can not be prepared from $\ket{X(n-1,S=(n-2)/2,M=(n-2)/2}$ by appending $ \ket{1}$. However if it was possible to prepare these states by these 'forbidden' paths, it would be of a great advantage as the state  $\ket{X(n-1,S=(n-2)/2,M=(n-2)/2}$ is completely un-entangled. This means that we could have prepared $\ket{X(n,S=n/2,M=(n-4)/2)}$ state in just two steps. The same considerations apply to preparation of $\ket{X(n,S=n/2,M=(-n+4)/2)}$ starting from $\ket{X(n,S=n/2-2,M=-n/2)}$ appending $\ket{0}$ states, which is a completely un-entangled state (product state).

To find a solution to this problem, let us take the case of preparing $\ket{X(n,S+1/2,M+1/2)}$ state from $\ket{\psi_i}=\ket{X(n-1,S,M+1)} \otimes \ket{1}$ and understand the constraint $2(S-2M)\geq 1$ from a different perspective. By using the equation \ref{eq:Ruben2}b, we can find out the average value spin angular momentum of the initial state as $\braket{\psi_i|\boldsymbol{S}^2|\psi_i}=\hbar^2[S(S+1)-M-1/4]$. The intermediate state ($\ket{\overline{\psi}}$ after the entangling evolution is given by equation \ref{eq:Ruben5}. If we can reach the intermediate state $\ket{\overline{\psi}}$, we can certainly reach the final desired state by correcting for the phase $\phi$ by single qubit operation. The average spin angular momentum of the intermediate state  can be found out using Eq.~\ref{eq:Ruben2}. After some algebraic steps, one can show that 
\begin{equation} \label{eq:spin_conserv}
\begin{aligned}
\braket{\overline{\psi}|\boldsymbol{S}^2|\overline{\psi}} &=\braket{\psi_i|\boldsymbol{S}^2|\psi_i}
\\ & + \hbar^2\frac{S+M+1}{2S+1} \left[2M+1+2(S-M)\cos(\phi)\right]
\end{aligned}
\end{equation} 
As the spin angular momentum is conserved during the evolution under all coupled Heisenberg interaction, the second therm on RHS of the above equation must be zero, which gives us a condition:
\begin{equation} \label{eq:phi_case1}
\cos(\phi)=1-\frac{1}{2}\frac{2S+1}{S+M+1}
\end{equation}
Comparing this to equation \ref{eq:ts_case4}, we see that the phase correction required is the same as $\omega t_s$. Further, the requirement that $\phi$ be real gives us the same condition as obtained before viz. $2(S-2M)\geq 1$. It should be noted that the spin angular momentum of the desired final state, $\ket{X(n,S+1/2,M+1/2)}$ given by $(S+1/2)(S+3/2)\hbar^2$, is always greater than that of the initial state. The difference, $(S+M+1)\hbar^2$  is actually supplied to the system during the phase correction step through single qubit operations which tends to make $\phi = 0$. Thus the algorithms that we have used essentially work as follows: The initial state is subjected to entangling evolution. Spin momentum  is conserved in this process. Then the state is subjected to z-axis rotation of last qubit. This step changes the spin momentum of the state. If the spin momentum after these two steps can reach the desired final state spin momentum, the expansion is successful. If the spin momentum of the final state can not reach the spin momentum of the desired final state, what we can do is to repeat the previous two steps i.e. subject the state to entangling evolution for a certain time and then z-axis rotation of the last qubit. We repeat these steps till the spin momentum of the evolved state reaches the final state spin momentum. Note that there is a limitation to how much spin momentum can be pumped using single qubit rotations (dependent on the current state) and hence we need to go through these processes iteratively to pump in more and more spin momentum. This redressal can be summarized as a schematic illustrated in Fig.~\ref{fig:AmplitudeAmplification} where we have simply appended a loop in the previous sequence of operations.

\begin{figure} [tbh]
\tikzstyle{decision} = [diamond, minimum width=2cm, minimum height=0.5cm, text centered, draw=black, fill=green!30, inner sep = -5pt]
\begin{tikzpicture}

	\node [left] at (0,0) (StartState) {$\ket{X(n-1,S,M+1)}$};
	\draw [->] (StartState.east) -- ++(0.7,0) node[above,pos = 0.5] (AddOne) {\scriptsize $\otimes \ket{1}$};
	\node [left] at (1.5,0) (InitState) {$\ket{\psi_i}$};
	\draw [->] (InitState.east) -- ++(0.7,0) node[above,pos = 0.5] (Entangle) {\scriptsize $U(t_s)$};
	\node [left] at (3.23,0) (EvolvedState) {$\ket{\overline{\psi}^{(r)}}$};
	\node [left] at (5.6,0) (FinalState) {$\ket{\psi^{(r)}}$};
	\draw [->] (EvolvedState.east) -- ++(1,0) node[above,pos = 0.7] (QubitRotation) {\scriptsize $I \otimes R_z(\theta)$} -- (FinalState);
    \node [decision] at (2.73,-1.8) (decide) {\tiny $E_{\psi^{(r)}} \stackrel{?}{<} E_{m}$};
    \draw [->] (FinalState.south) |- (decide.east);
    \draw [->] (decide.north) -| node[right,pos=0.5] {\scriptsize yes} 
        node [left, near end] {\scriptsize $U(t_s)$} (EvolvedState);
    \node [left] at (0.7,-1.8) (FinalState) {$\ket{X(n,S+1/2,M+1/2)}$};
    \draw [->] (decide.west) -- node[above,pos=0.1] {\scriptsize no}  ++(-1,0);    
\end{tikzpicture}
\caption[] {\textbf{Expansion using amplitude amplification in 'forbidden' regions}. We sequentially evolve (through $U$) the system for appropriate time ($t_s$) and pump in the maximum possible energy (through $R_z(\theta)$) to the evolved state $\ket{\psi^{(r)}(t_s)}$. We repeat this sequence with the phase corrected state ($\ket{\psi^{(r)}_f}$) as the initial state for evolution in the next iteration of the sequence. Here superscript $r$ counts the number of iterations. The sequence is repeated until the energy of the phase corrected state $\left( E_{\psi^{(r)}_f} \right)$ reaches ($E_m$) the energy corresponding to the desired Spin eigenstate $\ket{X(n,S+1/2,M+1/2)}$. Here, $I$ is the identity operation on the n-1 qubits.}
\label{fig:AmplitudeAmplification}

\end{figure}
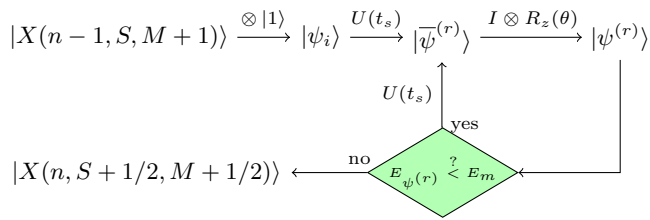

A rigorous way to appreciate the workings of this modification is to track the states throughout. Like seen before, tracking the coefficients $c_1(t), c_2(t), a_1(t)$ and $a_2(t)$ is sufficient for this purpose. It is also worth emphasising here that $(c_1(t), c_2(t))$ can be considered as a (complex) coordinate vector of the state at time t in the basis \{$\ket{X(n,S+1/2,M+1/2}$, $\ket{X(n,S-1/2,M+1/2}$\}. Studying time evolutions under all coupled Heisenberg interaction is easy in this basis. A state evolved for a time $t_s$ can be simply represented by the coordinate $\left(c_1(t) \exp(-i \omega t_s), c_2(t) \right)$ ignoring a global phase. Similarly, the coordinate $(a_1(t), a_2(t))$ in \{$\ket{X(n-1,S,M} \otimes \ket{0}$, $\ket{X(n-1,S,M+1} \otimes \ket{1}$\} basis is easy to work with, when studying the effect of single qubit rotation $R_z$ of the added qubit. A rotation by an amount $\theta$ transforms the coordinate to $\left(a_1(t) \exp(-i \theta), a_2(t)\right)$ where we have neglected the global phase again. Also, since we already have a way to transform the coordinate vectors into the other basis after either evolution or rotation operation (Eq.~\ref{eq:Ruben3}), we can find the state coordinates at any time during the iteration starting from the initial state $\ket{\psi_i}$ which is represented as $(c_1(0),c_2(0)) = (B,A)$ and $(a_1(0),a_2(0)) = (0,1)$ in respective bases. 

The choice of $t_s$ and $\phi$ in each iteration that minimizes the total number iterations required to prepare the desired state is an important question. We observe that except for the last iteration, choosing $t_s = \pi/\omega$ and $\theta = \pi$ takes the coefficients closest to the desired state coordinates $(c_1, c_2) = (1,0)$ and $(a_1,a_2) = (A,B)$ in respective bases for given values of S and M and hence A and B (See Appendix.~\ref{app:MapToGrover}). For this choice, the coefficients in the two bases at the end of $(r+1)$th iteration in terms of those at the end of $r$th iteration can be obtained as:
\begin{equation} \label{eq:ep}
\begin{aligned}
c_1[r+1] &= \left( A^2 - B^2 \right) c_1[r] +  2AB c_2[r] \\
c_2[r+1] &= -2AB c_1[r] + \left( A^2 - B^2 \right) c_2[r] \\
a_1[r+1] &= \left( A^2 - B^2 \right) a_1[r] +  2AB a_2[r] \\
a_2[r+1] &= -2AB a_1[r] + \left( A^2 - B^2 \right) a_2[r] 
\end{aligned}
\end{equation}
where we have discretely indexed the coefficients using square brackets indicating a sampling of their continuous counterparts at the end of each iteration. For r = 0 the sampling is done at t=0 i.e. $c_1[0]$ is defined as $c_1(0)$ and so on. Eq.~\ref{eq:ep} is suggestive of a rotation of the coordinate vector by same amount, say $\alpha$, in either basis and can be easily by choosing to write $(A,B) = (\cos(\alpha/2), \sin(\alpha/2))$. Further insight can be gained by visualizing these operations and hence the rotations geometrically and is discussed in detail in Appendix.~\ref{app:MapToGrover}. The key understanding is that the sequence of operations $(I \otimes R_z(\pi)) U(\pi/\omega)$, can be compared to a Grover's iterate, with the time evolution $U(\pi/\omega)$ seen as an instance of an oracle and the single qubit rotation $R_z(\pi)$ as an instance of the reflection operation about the initial state vector instead of the usual reflection about the mean operation (also known by the name diffusion operation). The initial iterations progresses identically to Grover's algorithm amplifying $c_1$, the component along $\ket{X(n,S+1/2,M+1/2)}$, in each iteration and hence pumping in the spin momentum described previously. This prepares the state to within an error probability of $\frac{S-M}{2S+1}$ in $O\left(\sqrt{\frac{2S+1}{S-M}}\right)$ iterations. However, this error can be corrected completely  in the last iteration. In the last iteration the evolution interval and the amount of rotation is chosen in a way discussed in previous section: Solving $|a_1(t_s)|=A$ (or $|a_2(t_s)|=B$) for $t_s$ after the entangling evolution and determining the relative phase between $a_1$ and $a_2$ immediately after the entangling evolution. The final iteration thus gives us the desired  state: $\ket{X(n,S+1/2,M+1/2)}$. 


The above amplitude amplification scheme generalizes the spin incrementing expansion of a $\ket{1}$ appended spin-eigenstate $\ket{n-1,S,M+1}$ outside the region $2(S-2M) \geq 1$ and reduces to the single iteration expansion inside the region as discussed in previous section. Although we have discussed here one of the four spin-incrementing and decrementing methods, it is also possible to generalize other expansion methods using the same procedure in respective 'forbidden' regions. 


\subsection{Perspective on Dicke states} \label{subsec:PerspOnDickeState}
Here, we summarize the expansion methods discussed above in the context of highly-entangled largest-spin-valued eigenstates (called Dicke states) and evaluate its cost of preparation from completely unentangled states. 

For a system of n qubits, in the notation we have used here, Dicke states can be written as $\ket{X(n,S=n/2,M=n/2-k)}$. The index $0 \leq k \leq n$ is called the hamming weight and for $k =0$ and $n$ correspond to unentangled spin-eigenstates that are trivial to prepare. We will also use a relatively common shorthand $D^n_k$ to denote these states from now on. We must use spin-incrementing path in preparing these states and the methods discussed in Section.~\ref{subsec:ExpansionMethod}, can be translated as

\vspace{5pt}
\noindent \textbf{Spin Incrementing Expansion Schemes}: 
\hrule \vspace{1pt} \hrule
\begin{enumerate}
  \item Weight preserving: \\
  $ D^n_k \xrightarrow[]{\otimes \ket{0}} 
  D^n_k \otimes \ket{0} \xrightarrow[]{U} \bar{D}^{n+1}_k \xrightarrow[]{\text{Phase Correction}} D^{n+1}_k $  \label{ES:es1}
  \item Weight incrementing: \\
  $D^n_k \xrightarrow[]{\otimes \ket{1}} 
  D^n_k \otimes \ket{1} \xrightarrow[]{U} \bar{D}^{n+1}_{k+1}\xrightarrow[]{\text{Phase Correction}} D^{n+1}_{k+1} $ \label{ES:es2}
\end{enumerate}
\hrule \vspace{1pt} \hrule 
\vspace{5pt}

For convenience we have chosen to write the expansion methods starting from an n qubit state instead of n-1 qubit state. We have also dubbed the expansion of a $\ket{0}$ and $\ket{1}$ appended eigenstate as weight preserving and incrementing expansion respectively. The respective requirements on $\widetilde{S}$ and $\widetilde{M}$ can also be translated in terms of n and k as $k \leq 3(n+1)/4$ and $k \geq (n-3)/4$ in respective cases. These expansions can be visualized in an n-k space where each coordinate $(n,k\leq n)$ represents a Dicke state $D^n_k$. Here, tracing points horizontally (diagonally) towards right (positive slope direction) corresponds to weight preserving (incrementing) expansion. Corresponding regions where aforementioned movement is feasible are shaded green and pink respectively in Fig.~\ref{fig:reachabiillty}. The dark coloured region indicates the simultaneous feasibility of two expansions and therefore it is possible to expand in either direction for points in this region. The blue coloured line borders the entangled states inside. Thus, the task of preparing an arbitrary Dicke state essentially translates to finding out a path from $(n,k)$ to any point outside the blue border representing all zeros or all ones states. It can be seen that the weight preserving and incrementing expansions are not sufficient individually to prepare arbitrary Dicke states but are sufficient when used together. As an example, consider preparation of $D^{34}_{10}$ state. As shown in  Fig.~\ref{fig:reachabiillty}, we back trace diagonally till point (31,7) and then horizontally until we reach (9,7). And finally we back trace diagonally to point (2,0). This is also evident from preparation of $D^{33}_7$ and $D^{35}_{28}$, whose preparation paths are also shown in Fig.~\ref{fig:reachabiillty}. Based on this observations, we can write a particularly simple linear-step algorithm to prepare arbitrary Dicke states in terms of finding a path to unentangled states: Back trace diagonally as far as possible, then back trace horizontally as far as possible and repeat these steps. 

In Section.~\ref{subsec:AmplitudeAmp} we discussed amplitude amplification can be used to expand eigenstates in the 'forbidden' regions. In the present context this means we can continue to trace a path diagonally backwards inside green region as well expecting to reach an unentangled state quicker. For the example of preparing $D^{34}_{10}$ considered above, we traced back a path to (31,7) but using amplitude amplification we can trace a path all the way back to (25,1) and hence (24,0) which is an simple product state $\ket{0}^{\otimes 24}$ (See Fig.~\ref{fig:reachabiillty}). Although, we reach an unentangled state in $k =10$ spin-incrementing steps here, we should note that each weight incrementing expansion (using amplitude amplification) in the green region becomes costlier, requiring more number of iterations (r), for decreasing k and therefore a sum of iterations in each step of the traced path is a better indicator of the cost. Also, since the cost of preparing $D^{n}_1$, which is a W state, from $D^{n-1}_0$ is $O(\sqrt{n})$, we may as well choose to prepare these W states exponentially faster using algorithms like discussed in Ref.~\citep{LogWstate} in $O(\log(n))$ steps. The cost for preparing $D^{n}_1$ from $D^{n-1}_0$ is obtained by noting that preparing $D^{n+1}_{k+1}$ from $D^n_k$ requires $O\left(\sqrt{\frac{2\widetilde{S}}{\widetilde{S}-\widetilde{M}}} \right) \approx O\left(\sqrt{\frac{n+1}{k+1}} \right)$ iterations, where we have substituted $\widetilde{S}=(n+1)/2$ and $\widetilde{M} = (n+1)/2-(k+1)$.

We can now describe our final algorithm to prepare arbitrary Dicke states.
For $k<n/2$, first prepare $D^{n-k+1}_1$ state in $\lceil \log_4 (n-k+1)\rceil$ number of stages as described in \cite{LogWstate}, then use the generalized weight incrementing method to successively prepare Dicke states with increasing "spin-down" number (climbing diagonally upwards in Fig.~\ref{fig:reachabiillty}). The number of stages required for each climb  is the number of convergence steps and is added up cumulatively with each climb upto the point (n,k). Note that r automatically becomes 1 when in pink region. The total time cost (number of stages) to prepare (n,k) can thus be given by:
\begin{equation} \label{eq:TotalCost}
\text{Cost(n,k)} = \lceil \log_4 (n-k+1)\rceil + \sum_{j=2}^{k} r_{(n-k+j,j)}
\end{equation}
where $r_{(p,q)}$ indicates number of iterations required to prepare $D^p_q$ from $D^{p-1}_{q-1}$. The exact cost in order to prepare $D^n_{k\leq n/2}$ using this scheme is plotted in Fig.~\ref{fig:TimeComplexity} for some values n and k. We see that the cost is always less than n and this algorithm performs significantly better than the linear step algorithm discussed previously for smaller and smaller values of k. The sub-linearity of this cost for some values of k is discussed in Appendix.~\ref{app:Sublinearity}. On the other hand, for $k>n/2$, prepare $D^{n}_{n-k}$ using the method above and then apply bit-flip gates to all the qubits. An analogous approach to prepare $D^n_{k > n/2}$ would to be to first prepare one-spin-up W states exponentially (also suggested in Ref.~\citep{LogWstate}) and then repeatedly use weight preserving expansion and if needed using amplitude amplification. 


As an example of using this final algorithm, a path to prepare $D^{34}_{10}$ state is shown in  Fig.~\ref{fig:reachabiillty}. Starting from point (1,1) we first jump by maximum factor of 4 to reach  point (25,1) i.e. $(1,1)\rightarrow (4,1)\rightarrow(16,1)\rightarrow(25,1)$. We then move diagonally till point (31,7) by using spin angular momentum pumping algorithm. From point (31,7) we move diagonally to point (34,10) by using weight incrementing method.   
   
\begin{figure}
\includegraphics[width=0.48\textwidth, trim={0 0 0 0},clip]{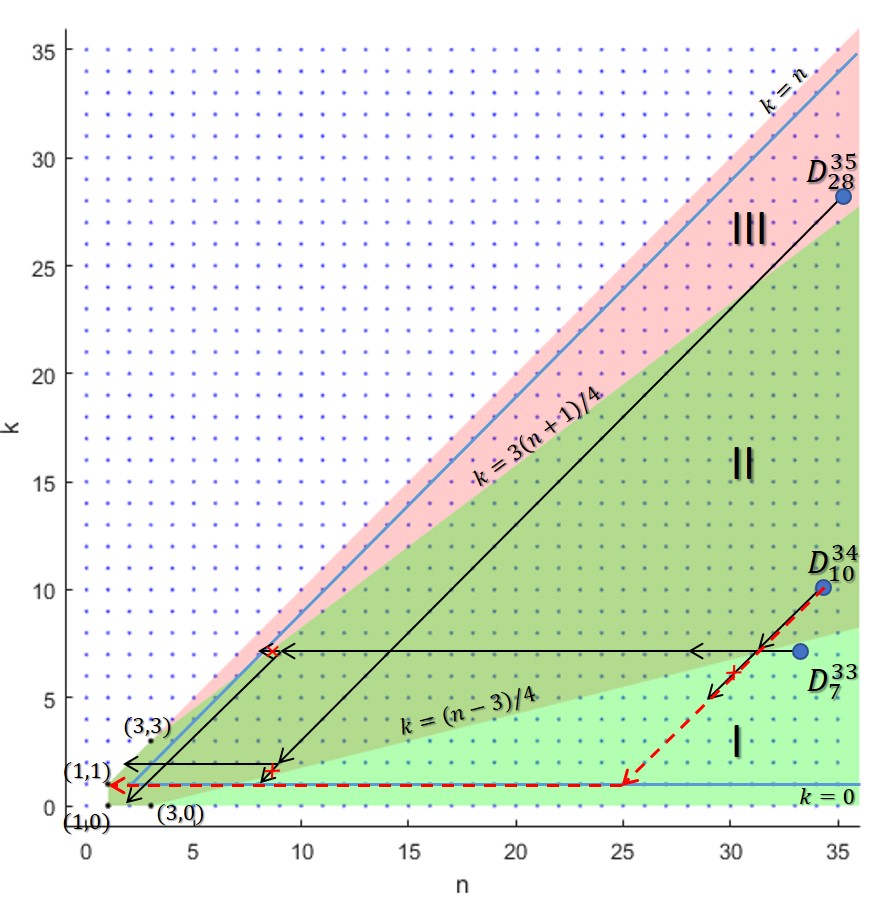}\caption[] {\textbf{Applicability of Dicke states expansion methods} The feasible choices of $n$ and $k$ for which weight incrementing and weight preserving expansion methods are marked in a plane described by the points $(n,k)$ as pink and green with respectively. There is an overlap in the two marked regions and is shaded darker and is labelled II. The exclusively green and pink regions are labelled I and III respectively. Three example preparation paths for preparing states in respective regions are shown and correspond to the linear algorithm. The path in dotted red corresponds to an example using the modified algorithm using amplitude amplification and logarithmic step preparation of W state.}
\label{fig:reachabiillty}
\end{figure}
   
\begin{figure}
\includegraphics[width=0.48\textwidth, trim={0 0 0 0},clip]{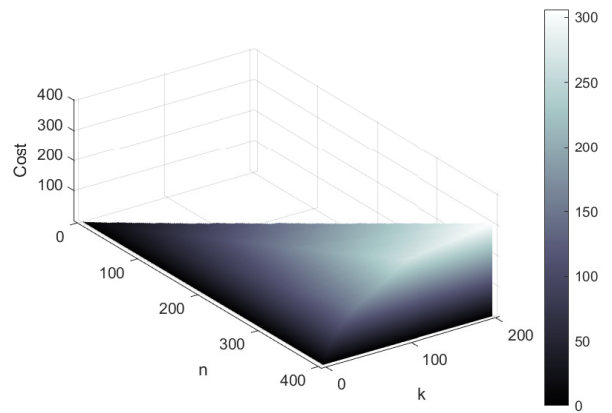}
\caption[] {\textbf{Cost of Preparation}. Total number of steps, consisting of entangling evolution and single qubit rotation, reguired to reach prepare $D^n_k$ are plotted as a function of $n$ and $k$ and correspond to numerical evaluations of Eq.~\ref{eq:TotalCost}.}
\label{fig:TimeComplexity}
\end{figure}


\subsection{Generalization of Hamiltonian}

We now discuss generalization of Hamiltonian describing systems where the algorithm discussed can still be implemented. The algorithms we have discussed used two kinds of Hamiltonians viz. Heisenberg exchange interaction between all pairs of qubits and Zeeman interaction generated by local magnetic fields along z direction acting on the qubits. Zeeman Hamiltonian is diagonal in the computational basis. All coupled Heisenberg Hamiltonian (Eq.~\ref{eq:Ideal_Hamiltonain1}) commute with $\boldsymbol{S_z}$ and is block diagonal in the partitions of the computational basis with fixed value of k ($k=n/2-M$). We will call this as k-spin-down subspace which has dimension of $^nC_k$. In this subspace, the Hamiltonian is given by:
\begin{equation} \label{eq:khotHamiltonian}
\mathcal{H}_k(i,j) = \bra{u_i} \mathcal{H}_k \ket{u_j} = 
\begin{cases}
      2J, & \text{if} \; \text{wt}(u_i \cdot u_j) = k-1\\
      0  & \text{otherwise}
\end{cases}
\end{equation}

where $\ket{u}$ denotes a computational basis state comprising of string of 0's and 1's. '$\cdot$' represents the bitwise 'AND'. Actually, every diagonal entry of $\mathcal{H}_k$ can be shown to be $J( ^k C_2 + ^{n-k} C_2 - k(n-k))$ but note we have taken them to be 0, since they only contribute global phase factors in state evolutions. (We have defined $J$ as $J^{\prime}(\hbar/2)^2$ for simplicity.) The above Hamiltonian has dimension of $^nC_k$ with $k+1$ distinct eigenvalues corresponding to S varying from $n/2$ to $n/2-k$. Equation \ref{eq:khotHamiltonian} essentially means that the Hamiltonian has  non-zero values at only the indices corresponding to the basis states obtainable by single swaps of 1 and 0. A more general Hamiltonian in the k-spin-down subspace can be given as ($k\leq n/2$)
\begin{equation} \label{eq:khotGeneralizedHamiltonian}
(\mathcal{H}_k)_{ij} = \bra{u_i} \mathcal{H}_k \ket{u_j} = 
\begin{cases}
	  2J_0, & \text{if} \; \text{wt}(u_i \cdot u_j) = k\\
      2J_1, & \text{if} \; \text{wt}(u_i \cdot u_j) = k-1\\
      2J_2, & \text{if} \; \text{wt}(u_i \cdot u_j) = k-2\\
			& \vdots \\
      2J_k, & \text{if} \; \text{wt}(u_i \cdot u_j) = 0\\
\end{cases}
\end{equation}

If we impose $J_m = 0 \; \forall \; m \neq l \in [0,k]$ above and denote the Hamiltonian (partial Hamiltonians) so obtained by $\mathcal{H}_{k,l}$, we can write $\mathcal{H}_k = \sum_{l = 0}^k \mathcal{H}_{k,l}$. The partial hamiltonian $\mathcal{H}_{k,l}$ can be interpreted to represent the interaction between the states obtainable by \textit{l} swaps of 1's and 0's in the binary literals representing these state. It can be seen that only $\mathcal{H}_{k,1}$ has been dealt with until now. The other partial Hamiltonians also have a similar symmetry as possessed by $\mathcal{H}_{k,1}$.
Therefore, the expansion procedures, scope of their operations and the amplitude amplification algorithm  work for each $\mathcal{H}_{k,l}$ independently ($l\neq 0,\mathcal{H}_{k,0}=I$). The only difference is that the value of $\omega =(E_1-E_2)/\hbar$  
depends on the $\mathcal{H}_{k,l}$ (See the discussion in Appendix.~\ref{app:GeneralHamiltonianExplanation}). Further, the partial Hamiltonians commute with each other i.e. $[\mathcal{H}_{k,m},\mathcal{H}_{k,n}] = 0$. Thus a Hamiltonian given by sum the partial Hamiltonians can also be used for all the algorithms discussed previously.     


\subsubsection{Special Case of all equal entries for exponential speedup}
In the specific case of $J_m = J \; \forall \; m \in [1,k]$ one may achieve exponential speedup in the number of expansion steps required for Dicke state preparation. Our previous work dealt with Hamiltonians with all equal non-diagonal entries and enabled exponential expansion (add ref).
If one can possibly engineer such a system with all equal entries in the k-spin-down Hamiltonian, then it would also be possible to prepare Dicke states in logarithmic number of stages. 

Consider again a Dicke state $D^n_k$. Suppose we add q number of qubits in state $\ket{0}$ so that the initial state is written as $D^n_k \otimes \ket{0}^{\otimes q}$. If we evolve it using the Hamiltonian of Eq.~\ref{eq:khotGeneralizedHamiltonian} with all equal $J_l$'s, the state at time t is given by: $\ket{\psi(t)}=a(t) D^n_k \otimes \ket{0}^{\otimes q}+b(t) \ket{\psi^`}$, where $\ket{\psi^`}$ is the sum of all computational basis states without q zeros in the end. The time evolution is stopped at time $t_s$ when $|a(t))|=|b(t))|$. The state time $t_s$ is denoted by $\bar{D}^{n+q}_{k}$.  Correcting the phases of last q qubits by z-axis rotation produces the $D^{n+q}_k$ state. Using the spin angular momentum conservation argument used before, we get the following condition on q for achieving $|a(t_s))|=|b(t_s))|$: 
\begin{equation}
\frac{\binom{n+q}{k}}{\binom{n}{k}} = 4 \sin^2(\frac{\phi}{2})
\end{equation}

where $e^{i\phi}=b(t_s)/a(t_s)$. We are interested in the maximum value of q satisfying this constraint.
One can similarly obtain a constraint on jumps in using weight incrementing method starting from $D^n_k \otimes \ket{1}^{\otimes q}$ evolving into $\bar{D}^{n+q}_{k+q}$. Given n and k, q must satisfy $\frac{\binom{n+q}{k+q}}{\binom{n}{k}} = 4 \sin^2(\frac{\phi}{2})$. Phase correction on the last q qubits of $\bar{D}^{n+q}_{k+q}$ gives us the desired $D^{n+q}_{k+q}$ state.


Note that in either case when $q_{max} = 0$, expansion is not allowed. It can be verified that $q_{max} = 0$ for points $(n,k)$ in regions outside green and pink regions for weight preserving and incrementing methods respectively. These regions infact defined the allowed regions for jump step of size 1. Clearly, larger jumps, $q>1$, can be made only while remaining these regions. To get an estimate on the speedup achieved in these cases, let's consider the factor by which the state expands in the limit of large n. We can call it an expansion factor: $ E.F. = \displaystyle \lim_{n \to \infty} (n+q_{max})/n $. It can be shown that $E.F. = 4^{1/k}$ in the weight preserving method for a given k. Since we move diagonally in the n-k space in the weight preserving method, it is beneficial to obtain this factor along one particular line say $k = n-p$, which gives us $E.F. = 4^{1/p} $. In either case E.F. is greater than 1 for a given k or p. This indicates an exponential speedup and hence it is possible to find a path to prepare arbitrary Dicke states in logarithmic number of steps. As an illustrative example consider the expansion of $D^{10}_4$ using the two methods while taking maximum possible jumps shown in Fig.~\ref{fig:maxJumps}. 


\begin{figure}[tbp]
\includegraphics[width=0.48\textwidth, trim={0 0 0 0},clip]{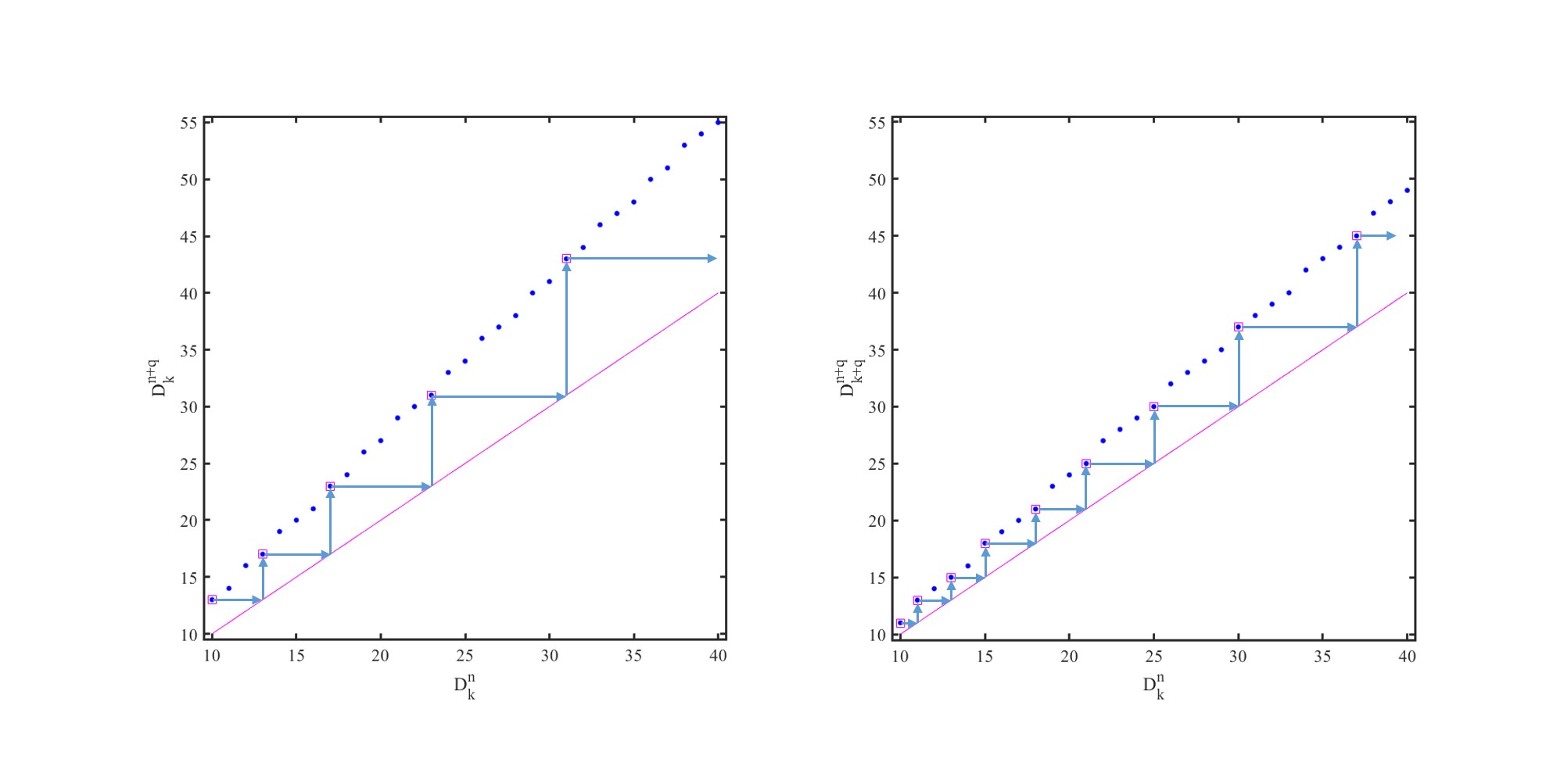} 
\caption [] {\textbf{Jumps}. Expanding from state $D^{10}_4$ while taking maximum possible jumps using (a) weight preserving strategy and (b) weight incrementing strategy.}
\label{fig:maxJumps}
\end{figure}



\subsubsection{Caveats on choice of relative strengths}
While the feasibility of the exapansion method for a Dicke state is dependent upon the symmetry of the Hamiltonian and initial state at hand, the speed of evolving it to the desired state is in general dependent upon the strengths of interactions (the values of $J_l$'s.). The relative strengths may increase or decrease the speed of the evolution. However in certain extreme cases, the evolution can be completely killed despite the symmetry. It can be noted 
that the energy difference $E_1-E_2$ influences the time required for entangling evolution. It can happen that for certain ratios of $J_l$'s in  Eq.~\ref{eq:khotGeneralizedHamiltonian},  $E_1-E_2=0$ i.e. evolution will take infinite time or in other words evolution does not take place. This essentially means that the starting state is an eigenfunction of the Hamiltonian and evolution simply gives a phase factor. Therefore, care must be taken while engineering this Hamiltonian.

As an example consider preparation of state $D^n_2$. If the Hamiltonian of the system is given by $\mathcal{H}_2= \mathcal{H}_{2,1} + \mathcal{H}_{2,2}$, the factor $E_1-E_2$ is given by $2 J_1 n+ J_2 n(n-3)$. (See the last paragraph of appendix \ref{app:GeneralHamiltonianExplanation} for the energy factor with $J_2$ coefficient.) Clearly if $J_1/J_2=(3-n)/2$, the energy difference is zero. Thus the state $D^n_2$ can not be reached from $D^{n-1}_2\otimes \ket{0} $ state by using this Hamiltonian.


%
%
%
%
%
%
%

\section{\label{sec:sTT} Use case scenario: Spin based Quantum Computing architecture where information is written through spin torques}

\begin{figure}[b]
\includegraphics[width=0.48\textwidth, trim={0 0 0 0},clip]{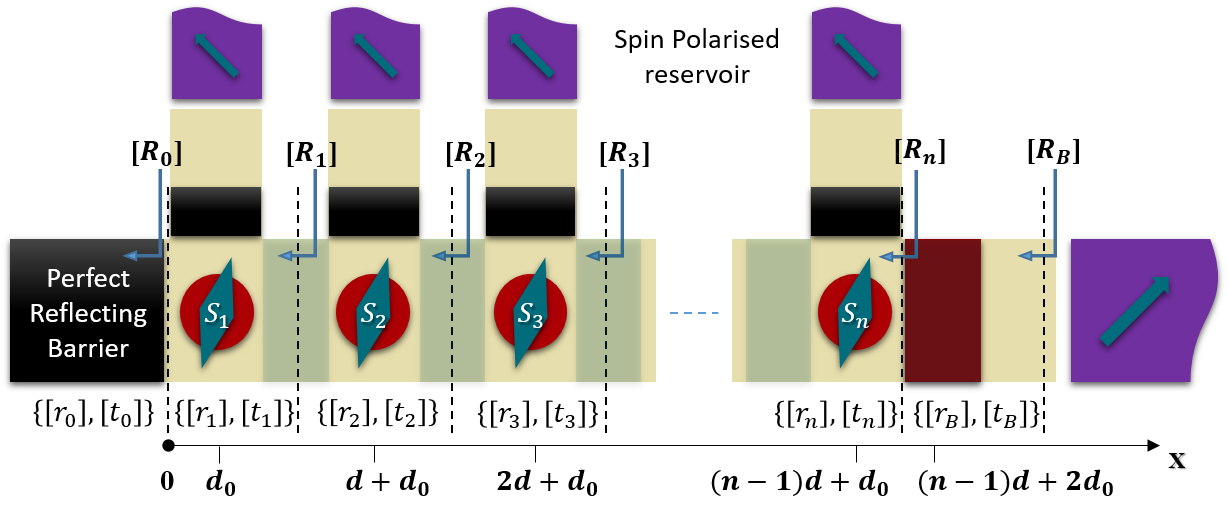}
\caption[]{\textbf{Schematic of the system} (\textit{Figure reproduced from} \citep{LogWstate}). n static qubits (colored red) in a spin coherent channel (shaded yellow). There are barrier gates (colored black) to facilitate creation of standing waves and a reservoir (colored purple) to inject and extract spin polarized carriers. The distance between two successive qubits is $d$ while that between a qubit and a barrier gate is $d_0$. Individual qubits act as spin-dependent scatterers with reflection and transmission denoted by $[r,t]$ matrices. Reflection matrices looking into the cascade of scatterers is also shown.}
\label{fig:QubitInChannel}
\end{figure}

Consider a collection of spin-impurities arranged in a spin-coherent medium with no mutual interaction. A spin-polarized reservoir injects electrons into the system that interact with individual impurities via Heisenberg exchange. (Repeated) Scattering of injected electrons (flying qubits) from (successive) impurities mediates an effective interaction between the impurities (static qubits) that potentially can entangle their spin states. Hard barriers on the periphery completely reflect these electrons and they are eventually ejected back to the reservoir after multiple scattering from the impurities and additional barriers (controlled electrically) placed in the medium.  

With proper design, the eventual reflection of the flying qubit from medium back towards the reservoir can be associated with a unitary operator $R_B$ in the combined hilbert space of flying and static qubits. Note that hard barriers and hence complete reflection towards the reservoir is necessary for unitarity of $R_B$ \citep{Ciccarello_2012}. If we denote the combined initial state of the injected flying qubit and the n-qubit system as $(\rho_f \otimes \rho_s)$, where $\rho_f$ and $\rho_s$ are $2 \times 2$ and $2^n \times 2^n$ density matrices respectively, then the reflection process characterised by $R_B$ modifies the state to $R_B (\rho_f \otimes \rho_s) R_B^{\dagger} $. Extraction of flying qubits back to reservoir causes the state to collapse, therefore, the modified state of static qubits can be written as $\text{Tr}_f \left[ \mathcal{R}_B (\rho_f \otimes \rho_s[m]) \mathcal{R}_B^\dagger \right]$, where Tr$_f$ denotes partial trace over the flying qubit. This transformation of $\rho_s$ is considered as one evolution step. We repeat the process with the modified state for every newly injected electron which can be controlled through use of barriers. The scheme requires several such injection-interaction-extraction cycles and therefore the state of static qubits system can be indexed by number of injected electrons (N). The evolution step described above can be rephrased in the language of Kraus operators \{$M_k$\} satisfying $\sum_k M_k^{\dagger} M_k = \mathcal{I}$ so that the evolved state after electron extraction can as well be written as $\sum_k M_k \rho_s M_k^{\dagger}$.

We have studied the above system in our previous work to illustrate preparation of W-states in logarithmic number of steps. We briefly highlight the procedure to construct $R_B$ below.  The specific example we have considered (and also consider here) is a non-interacting chain of spin-1/2s in 1D. The barriers and static qubits are increasingly labelled towards right as shown in Fig.~\ref{fig:QubitInChannel}. Being localized in space, they are considered as delta potential scatterers. Scattering from spin qubits is accounted for spin-dependence by assuming an exchange interaction between flying and static qubit. Transmission and reflection from the $j^{th}$ scatterer are thus described by following matrices,
\begin{equation}
t_j =
\begin{cases}
	  [\mathcal{I} + i \Omega \bm{\sigma_f} \cdot \bm{\sigma_j}]^{-1} & \text{for static qubits} \\
	  [\mathcal{I} + i \Gamma \mathcal{I}]^{-1} 
	  				& \text{for additional barriers}\\
\end{cases}
\end{equation}
and $r_j = t_j - \mathcal{I}$. Here, $\Omega$ and $\Gamma$ are parameters proportional  to respective barrier strengths, $\sigma_f$ and $\sigma_j$ correspond to spin operators of flying and $j^{th}$ static qubit respectively and $\mathcal{I}$ is $2^{n+1} \times 2^{n+1}$ identity matrix.

The overall reflection matrix $R_B$ thus can be constructed by cascading reflection matrices iteratively as follows:
\begin{multline} \label{eq:RefMatCascade}
\hat{r}_j = 
\begin{cases}
r_0 & \text{if} \; j = 0 \\
r_j + \alpha t_j \left(\mathcal{I} - \alpha \hat{r}_{j-1} r_j \right)^{-1} \hat{r}_{j-1} t_j & else
\end{cases}
\end{multline}
with $\alpha := e^{2i kd_j}$ where $d_j$ is the distance from the previous scatterer and k is the wave-number of injected electrons. Note $R_B = \hat{r}_B$ in accordance with Fig.~\ref{fig:QubitInChannel} and above equation. A hard barrier with $\Gamma \rightarrow \infty$ i.e. $t_0 = 0$ at the left end in Fig.~\ref{fig:QubitInChannel} ensures $R_B$ is a unitary operation enabling to perform quantum evolutions as descirbed previously. 
Although we can use either polarization to demonstrate the scheme, we will base the following discussion on injection of $\ket{0}$ polarized electrons from the reservoir. We can express $\mathcal{R}_B$ in an (n+1) qubit computational basis as a matrix and partition it into four matrix blocks each of size $2^n \times 2^n$. The relevant Kraus operators $M_0$ and $M_1$ are thus given by the top left and bottom left blocks respectively.
satisfying $\text{M}_0^\dagger \text{M}_0 + \text{M}_1^\dagger \text{M}_1 = \mathcal{I}_{2^n}$. 

It turns out that the matrix elements of $R_B$ or for that matter $M_0$ and $M_1$ are a function of the four parameters of the system: $kd$, $kd_0$, $\Gamma$ and $\Omega$. With appropriate choice of parameters $M_1$ can be made close to 0 so that the evolved state can be almost unitarily evolved using $M_0$ alone as $M_0 \rho_s M_0^{\dagger}$. For the purposes of demonstration of the expansion schemes (and hence the modified algorithm), we choose these paratmeter values as $(kd,kd_0) = (\pi,\pi/2)$ and $(\Gamma,\Omega) = (1000,0.0001)$, which were optimized for three-qubit one-spin-down subspace in previous work \citep{LogWstate} and also turn out to be good enough parameters for other k-spin-down subspaces. 
With these parameters, $M_0$ acceptably emulates the unitary corresponding to the Hamiltonian given by Eq.~\ref{eq:khotGeneralizedHamiltonian} in a given subspace. For a small interval $\delta t$, the corresponding unitary evolution $ U_k(\delta t)$ can be written as $ U_k (\delta t) = \mathcal{I} - i \mathcal{H}_k \delta t $ and corresponding strength parameters $J_0, J_1, J_2$, etc can be extracted from the $R_B$ matrix description. Note that there are no apriori assumptions that force $J_m$s may to be constant across subspaces nor with the number of qubits in the system. Fig.~\ref{fig:J_3hot_M0} shows effective strengths $J_m \delta t$ of $M_0$ in the 3 spin-down subspaces of n-qubits obtained from $-\operatorname{\mathbb{I}m}\{U_k(\delta t)\}/2$.

\begin{figure}
\centering
\includegraphics[width=0.48\textwidth, trim={0 0 0 0},clip]{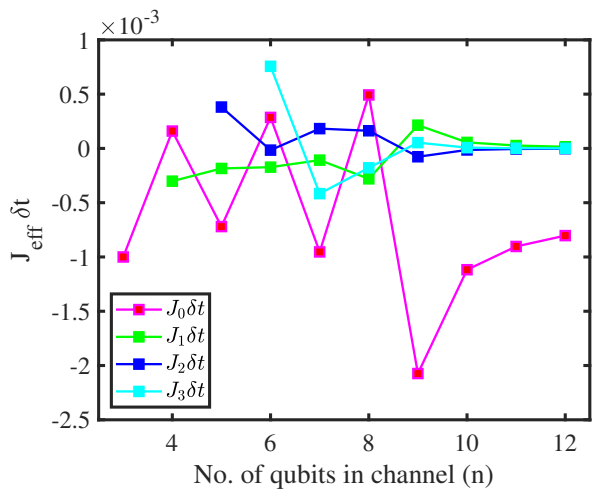}
\caption[] {\textbf{Interaction Strengths as a function of number of qubits} $J_0 \delta t$, $J_1 \delta t$, $J_2 \delta t$, $J_3 \delta t$ in three-spin-down subspace.
}
\label{fig:J_3hot_M0}
\end{figure}



We will now illustrate the expansion of $D^5_2$ state using the two expansion methods highlighted in Section \ref{subsec:PerspOnDickeState} in this spin torque quantum computing architechture. Assume there are six qubits arranged in channel like shown in Fig.~\ref{fig:QubitInChannel}. Say the left five qubits are entangled in state $D^5_2$. For the weight preserving method the sixth qubit 
is arranged in state $\ket{0}$ while it is arranged to be in state $\ket{1}$ for the weight incrementing method. These single qubit states in this architecture can be prepared by connecting them directly to the desired polarized reservoir for long time \citep{sutton2015manipulating}. One may as well inter-convert $\ket{0}$ to $\ket{1}$, if the former state is already available, or vice-versa, through single qubit rotation say about y-axis (rotation about z-axis is explained later).
Now, to entangle all qubits with each other, all the barriers between the qubits used for isolating are lowered for zero reflection ($\Gamma = 0$) while the barrier next to the sixth qubit before the reservoir (colored burnt umber) is electrically lowered for partial transmission ($\Gamma \neq 0$). This allows polarized electrons to enter the channel and eventually get ejected after interaction as described previously. During this phase the state of the 6 qubits evolves with two kinds of entries on the diagonal of its density matrix which are proportional to $|a_1(t)|^2$ and $|a_2(t)|^2$ as described in section \ref{subsec:ExpansionMethod}. Since the role of time is taken up by number of electrons (N) injected or for that matter ejected from the channel, it is more suited to call them $|a_1(N)|^2$ and $|a_2(N)|^2$ and their corresponding scaled versions in computational basis as $|d_1(N)^2$ and $|d_2(N)^2$. Their evolution is shown in Fig.~\ref{fig:SixQubitEvolution}. At one point ($N = N_s$) the two curves cross each other ($N_s$ can be considered to be similar to $t_s$). We raise all barrier gates for zero transmission as close to the ideal intersection point as possible (given the evolution is discrete with number of electrons injected), shutting off the inflow of further electrons as well as isolating the neighbouring qubits which stops further evolution. 
An estimate for the number of electrons can be obtained using Eq.~\ref{eq:ts_case1} or Eq.~\ref{eq:ts_case3} plugging in appropriate value of $\omega$ estimated using the values of $J_m$s obtained from $-\operatorname{\mathbb{I}m}\{U_k(\delta t)\}/2$ (as explained previously). This may require further calibration in number of electrons in physical realization for better fidelities. In general if the changes in $|a_1(N)|^2$ or $|d_2(N)|^2$ per electron count is small, then better calibration is expected. At this stage a $\bar{D}^{6}_2$ or $\bar{D}^{6}_{3}$ state is formed depending on the initial state of the sixth qubit. Now for the phase correction we need to perform single qubit rotation of the 6th qubit about z-axis. For this, since the hard barriers are raised everywhere isolating the qubits, we lower only the barrier connecting the 6th qubit to a reservoir especially connected to it, so that electrons injected only interact with the 6th qubit (The hard barriers around the qubit also avoids any leakage towards other qubits in channel). Such a method of single qubit rotation is described in \citep{sutton2015manipulating}. Note that z-polarized reservoirs (the specific polarization only changes the sense of rotation) Again after certain amount of electrons have interacted calibrated for maximum fidelity between the current state of 6 qubit system with the expected state, we shut the gates even from the reservoir's side to completely cut off the qubit system from the environment. Fidelity is one way to determine the closeness of two states. For any state $\rho$ and another pure state $\ket{\psi}$, we use $F(\rho, \ket{\psi}) = \sqrt{\bra{\psi} \rho \ket{\psi}}$ as the definition of Fidelity \citep{nielsen2010quantum}. We obtain $\sim 99.9 \%$ fidelity for $D^6_2$ and $D^6_3$ prepared using the methods outlined above.


\begin{figure}
\centering
	\includegraphics[width=0.45\textwidth, trim={0 0 0 0},clip]{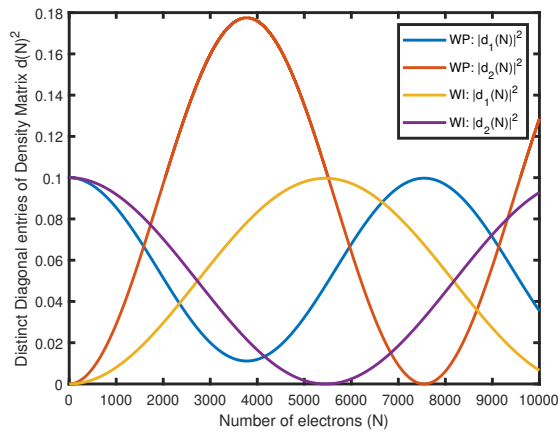}
\caption[] {\textbf{Six Qubit Evolution}. Evolution of distinct diagonal entries of system's density matrix represented in the 2-spin-down and 3-spin-down subspaces of the 6 qubit Hilbert space in respective expansion schemes. Here, WP and WI stand for 'weight preserving' and 'weight incrementing' respectively. Note that the y axis denotes the actual square amplitudes of coefficients and are proportional to respective $a_1$ and $a_2$ coefficients.}
\label{fig:SixQubitEvolution}
\end{figure}


Table \ref{tab:WeightPreservingFids} and \ref{tab:WeightIncrementingFids} summarizes the Fidelities obtained using these expansion procedures starting from a pure dicke states. For certain starting states like $D^9_1$ in the weight incrementing method the fidelities obtained are not great which is not surprising since it lies outside the feasible region obtained analytically shaded pink in Fig.~\ref{fig:reachabiillty}. Note the fidelities reported are after the phase correction step. For the cases like expansion of $D^9_1$, $|d_1(N)|^2$ and $|d_2(N)|^2$ never crossed and we stopped the entangling evolution when the two curves came closest and then performed phase correcting single qubit rotation. Except for such outliers the fidelities obtained are all $\geq 99\%$ when starting state parameters n and k lie in feasible regions for the expansion steps to work. For such outliers repeated pumping of spin angular momentum, as explained in Section.~\ref{subsec:AmplitudeAmp} is the correct method to follow as reflected in the improved the fidelities obtained by following this method. The improved fidelities are also reported in corresponding tables within paranthesis. 

Above, the starting Dicke states were considered to be absolutely pure and only reflect the efficiency of the expansion methods. More appropriate fidelity numbers can be obtained by utilizing the smaller sized Dicke states themselves obtained using either expansion methods. As a specific example, the following series of expansion steps as prescribed in the previous section without energy pumping: $\ket{11} \rightarrow D^3_2 \rightarrow D^4_2 \cdots \rightarrow D^9_2 \rightarrow D^{10}_3$ prepares $D^{10}_3$ with 98.43\% fidelity.
The phase correction step is a part of the jumps and is performed after each entangling evolution with the appended qubit and the fidelity reported is inclusive of the errors accumulated in each step. We also suggested an improvement to reduce the cost by first preparing W states exponenetially and using the generalized expansion method (angular momentum pumping) if required to prepare Dicke states with non-singular excitations. Table \ref{tab:FidelityModifiedAlgorithm} summarizes the Fidelities obtained for various Dicke states obtained using the improved algorithm. 
In reporting the fidelities we have assumed the initial products states of all zeros or all ones are available with $100\%$ fidelity. The qubits are entangled by allowing the injected electrons to interact with all spin impurities in channel and single qubit rotations are performed by allowing them to interact with only the target qubits by appropriate control of barrier gates as explained previously.




In the examples for which we have reported values, we have only considered $\ket{0}$ electrons from the reservoirs. But the following general rule of thumb would give better fidelities: Use $\ket{0}$ polarized electrons for $k<(n+1)/2$ and $\ket{1}$ polarized electrons otherwise, where n is the number of qubits in the channel considered for entangling evolution in pursuit of preparing $\bar{D}^n_k$. One can understand this by again looking at the matrix description of $R_B$ in computational basis which also happens to be block diagonal like the all coupled Hamiltonian. We can partition its $\binom{n+1}{k}$ dimensional basis (k spins down subspace of $n+1$ qubit Hilbert space) corresponding to flying qubit being in $\ket{0}$ or $\ket{1}$. There would be $\binom{n}{k}$ and $\binom{n}{k-1}$ basis states respectively in such partitions. One can argue in this k-spin-down subspace of combined Hilbert space of flying and static qubits, evolutions are closed. So any initial superposition are shared among all basis states in general. We desire final combined evolved state before taking a partial trace to look like $\ket{0} \otimes D^n_k$. Thus we want minimal sharing of superpositions with basis states corresponding to flying qubits in state $\ket{1}$. This can be translated to $\binom{n}{k} > \binom{n}{k-1}$ which is true for $k<(n+1)/2$.

For the simulation purposes we have used $\mathcal{R}_B$ matrices. The discussion on Kraus operators are just to highlight that they are used for optimization. And also because albeit with whatever optimization and design a full fledged simulation is more close to the physically realizable system ofcourse within the one dimensional assumption. Also, note that there are more than just two unique coefficients $d_1$ and $d_2$ in these evolutions (due to slight imperfections in emulating the hamiltonian and perhaps numerical errors) but can be classified just in those two representative groups broadly. The exact location when the simulations are stopped correspond to the point when the product of desired diagonal entries correspond to a maximum. Also for the phase correction part we stop the simulation whenever the fidelity reaches a maximum which can be shown to correspond injectively (one-to-one) to maximum energy possible for that iteration.


\begin{table}[tbp]
\caption{\label{tab:WeightPreservingFids}
Fidelities of states $D^{n+1}_k$ obtained after phase correction procedure in the weight preserving method starting from a $D^n_k$ state}
\begin{ruledtabular}
\begin{tabular}{ccccccccccccc}
$n \downarrow \ k \rightarrow $ & 1 & 2 & 3 & 4 & 5 & 6 & 7  \\
\hline
2  & 99.89 &       &  	   &   	   &       &  &               \\
3  & 99.95 & 99.85 &       &   	   &       &  &               \\
4  & 99.99 & 99.92 & 99.75 &   	   &       &  &               \\
5  & 99.99 & 99.96 & 99.89 & 99.72 &       &  &               \\
6  & 99.99 & 99.96 & 99.92 & 99.83 & 99.52 &  &                \\
7  & 99.99 & 99.97 & 99.94 & 99.88 & 99.74 &  98.46 &         \\
8  & 99.99 & 99.97 & 99.95 & 99.9  & 99.82 &  99.53   & 97.88  \\
9  & 99.99 & 99.97 & 99.95 & 99.92 & 99.86 &  95.17   & 98.51  \\
\end{tabular}
\end{ruledtabular}
\end{table}

\begin{table}[tbp]
\caption{\label{tab:WeightIncrementingFids}
Fidelities of states $D^{n+1}_{k+1}$ obtained after phase correction procedure in the weight incrementing method starting from a $D^n_k$ state. Entries in bracket indicate energy pumping.}
\begin{ruledtabular}
\begin{tabular}{ccccccccccccc}
$n \downarrow \ k \rightarrow $ & 1 & 2 & 3 & 4 & 5 & 6 & 7  \\
\hline
2 & 99.87 &       &       &       &       &   &           \\
3 & 99.92 & 99.91 &       &       &       &   &           \\
4 & 99.83 & 99.77 & 99.74 &       &       &   &           \\
5 & 99.88 & 99.83 & 99.77 & 99.6  &       &   &           \\
6 & 99.9  & 99.85 & 99.8  & 99.71 & 99.7  &   &           \\
7 & 99.91 & 99.86 & 99.82 & 99.75 & 99.39 & 99.87 &       \\
8 & 99.43 (99.83) & 99.87 & 99.83 & 99.77 & 99.65 & 99.84 & 99.9     \\
9 & 98.31 (99.84) & 99.87 & 99.84 & 99.8  & 99.71 & 99.2 & 99.8  \\
\end{tabular}
\end{ruledtabular}
\end{table}

\begin{table}[tbp]
\caption{\label{tab:FidelityModifiedAlgorithm}
Fidelities of states $D^{n}_k$ obtained using the modified Algorithm (using amplitude amplification)}
\begin{ruledtabular}
\begin{tabular}{ccccccccccccc}
$n \downarrow \ k \rightarrow $ & 1 & 2 & 3 & 4 & 5 & 6 & 7 & 8 &  \\
\hline
2  & 99.89 &   	   & 	   &   	   &       &       &       &       \\
3  & 99.85 & 99.68 &   	   &   	   &   	   &   	   &       &       \\
4  & 99.9  & 99.77 & 99.68 &       &   	   &   	   &   	   &       \\
5  & 99.89 & 99.73 & 99.43 & 99.63 &   	   &   	   &       &       \\
6  & 99.88 & 99.77 & 99.62 & 99.42 & 99.59 &       &       &       \\
7  & 99.87 & 99.78 & 99.66 & 99.23 & 99.38 & 99.54 &  	   &       \\
8  & 99.86 & 99.78 & 99.65 & 99.54 & 99.16 & 99.34 & 99.5  &   	   \\
9  & 99.86 & 99.70 & 99.66 & 99.52 & 98.97 & 99.13 & 99.22 & 99.47 \\
\end{tabular}
\end{ruledtabular}
\end{table}

\section{\label{sec:conclusion} Discussion and Conclusion}

In this article, we have examined a deterministic scheme to expand a given spin-eigenstate in an all to all symmetrically coupled system of spin qubits. We describe how addition of a single qubit in ground or excited state to a given n-1 qubit spin eigenstate can yield another n qubit spin eigenstate approximately using unitary time evolution of the combined n qubit system in four ways. The obtained state is only approximate and can be corrected 
using single qubit rotations. 
These expansions do not work universally for all S and M values but
a combination of these expansion methods together are sufficient to prepare arbitrary spin eigenstates 
in linear time. Next, an idea is proposed to improve this cost starting from product states or W states in conjugation with previous methods
and necessitates an expansion outside feasible choices. 
There is an energy gap between initial and final states and is not compensated by single qubit rotations for states outside feasible regions. Repeatedly pumping energy maximally using these rotations sandwiched with entangling evolution solves this problem. A comparison with the Grover's search is also presented.
For the special case of Dicke states, with specific number of down-spins we show that our algorithm is sub-linear in cost of preparation and otherwise is always better than the linear cost. This is followed by a discussion of generalized exchanged coupled interaction of all qubits where also our scheme can be implemented. Exponential speed-ups from the linear cost that can be achieved if certain parameters in the generalized Hamiltonian can be engineered desirably.

Since the all to all connectivity is a major hindrance to scalability, in the next segment of this paper we consider a spintronic quantum computing architecture based on static and flying qubit interaction suitable for universal fault-tolerant quantum computation, where an all to all connectivity can be indirectly realized. The evolution 
in this architecture is closely related to the idea of weak measurements in a system coupled to an ancilla. Here, flying qubit is the ancilla system that interacts with the (not mutually interacting) static qubits (spin impurities in a spin-coherent channel) which is the primary system successively and the operator to transform the state of ancilla coupled primary system is obtained using scattering theory. The flying qubit is eventually extracted (by a reservoir) which can be associated with an act of (projectively) measuring the flying qubit without post-selection. The post-measurement state of the primary system is obtained by tracing out ancilla from the combined transformed state. Under suitable design, it is possible to engineer the kraus operators to emulate the unitary corresponding to the all-coupled hamiltonian in appropriate subspaces. It turns out that this actually emulates the unitary corresponding to generalized all-coupled Hamiltonian whose parameters $J$s can be obtained on aprropriate comparison with appropriate the Kraus Operators. Therefore, using an ancilla one can indirectly achieve an all-to-all coupling in a system of otherwise non-interacting qubits and hence serves as an excellent use case for demonstrating our scheme. 


For the present article we chose to demonstrate the specific case of Dicke states preparation and obtained high quality states (Fidelity ~99\%) of upto 9 qubits and 8 excitations (c.f. Table.~\ref{tab:FidelityModifiedAlgorithm}) in MATLAB simulations. The parameters, both geometrical ($kd$ and $kd_0$) and interaction strength ($\Gamma$ and $\Omega$), were optimized for 3 qubits and 1-spin-down subspace. There is a scope for further improvement if the design is optimized for evolution in the intended subspaces. Ability to tune these parameters in real time can be also beneficial to improve Fidelity (using appropriate techniques). For example, the use of tunable resonant tunnelling barriers just before the reservoirs to select electrons with suitable values of k while the parameter $\Gamma$ can be controlled electronically. 

There are a number of considerations involved in physical design of this architecture and is similar to the discussion in \citep{sutton2015manipulating, LogWstate}. The non-idealities in physical design therefore affect the algorithm implementation. Some challenges in the physical implementation are as follows: Completely polarized reservoirs are rare in practice. But with it is possible to use reservoirs with polarizations as low as 30\% to obtain states with manageable loss in Fidelity. Another assumption of this model is availability of spin coherent channels. It is possible to achieve spin coherence lengths on the scales of micrometers \citep{Huang2007} but may become a challenge for large number of qubits. Scaling up the system size may require lower temperatures which has an added benefit of sharpening the peak value of wave-number k at which the electrons are injected. This is another way to address the assumption of monochromatic electrons (single value of k) of this model. The physical design can become even more complex with readout apparatuses and necessitates further optimization in the design. Also, it is assumed that an electron is injected only after an electron injected in the previous iteration is extracted. But we believe it should be possible to achieve such a control through passage of pulses of low spin currents in theory. The most important assumption of this model is that the spatial part of the interaction of flying qubits with the static qubits is assumed to be delta functions. Deviations from this ideal case may lead to additional complications and such an analysis is beyond the scope of the current paper.



Looking forward, as far as the algorithm is concerned, we understand that this preparation scheme should be implementable in systems where such an all to all connectivity can be directly or indirectly engineered. For three qubits, preparation of W state, a special case of Dicke state) has been experimentally demonstrated in a superconducting circuit QED system, where the constituent qubits were coupled directly with each other in a similar fashion \citep{Neeley_2010}. For the indirectly engineered all coupled system, experimental demonstration of the single qubit rotations, reliant on the scattering based model used here, should be first sought for. 



\begin{acknowledgments}
We acknowledge the support of Department of Science and Technology (DST), Government of India through Project No. SR/NM/NS-1112/2016 and Science and Engineering Research Board (SERB) through Project No. EMR/2016/007131.

\end{acknowledgments}

\bibliography{apssamp}


\appendix

\section{Extended Discussion on Amplitude Amplification method using Spin Pumping} \label{app:MapToGrover}

\subsection{On the choice of $t_s$ and $\theta$}

We have justified that the coefficients $c_1, c_2, a_1, a_2$ are sufficient to track the state at any point of time (in the iterative procedure). Let's consider the state just after the entangling evolution is performed described by complex coordinate $(a_1(t_s), a_2(t_s))$. Here, for notational simplicity we assume $t_s$ is counted from the beginning of the $r^{th}$ (current) iteration. Without loss of generality we can the write the state as (much like Eq.~\ref{eq:Ruben5}) 
\begin{equation} 
\begin{aligned}
\ket{\psi(t_s)} &= \left|a_1(t_s)\right| \exp(-i\phi) X(N-1,S,M)\otimes \ket{0} \\ 
 & + \left|a_2(t_s)\right| X(N-1,S,M+1) \otimes \ket{1}
\end{aligned}
\end{equation}
We can obtain the expectation value of $\braket{S^2(t_s)} = \braket{\psi(t_s)|S^2|\psi(t_s)}$ corresponding to above state as
\begin{equation} 
\begin{aligned}
\braket{S^2(t_s)} &= S(S+1/2) - (M+1/4) + (2M+1) \left|a_1(t_s)\right|^2 \\ 
 &+ 2(2S+1)AB\left|a_1(t_s) a_2(t_s)\right| \cos(\phi)
\end{aligned}
\end{equation}
The state after $R_z(\theta)$ and corresponding expected $S^2$ value can be obtained from above equations simply by substituting $\phi + \theta$ for $\phi$. We can see that the Spin momentum is pumped maximally if $\theta=2m\pi-\phi$ for any integer $m$. This settles the choice for $\theta$ given a particular state coordinate $(a_1, a_2)$ just before single qubit rotation. 

Now we will determine $t_s$ that maximizes $\braket{S^2}$ after the single qubit rotation. 
For this let's start by writing $a_1(t_s)$ and $a_1(t_s)$ in terms of the initial state coefficients as
\begin{equation} \label{eq:TimeAngleProof1}
\begin{aligned}
a_1(t_s)=A c_1[r-1] \exp(-i\omega t_s)-B c_2[r-1]\\
a_2(t_s)=B c_1[r-1] \exp(-i\omega t_s)+A c_2[r-1]
\end{aligned}
\end{equation}
where we have accounted for the entangling time evolution. For the following we will loose the r-1 index and $(c_1,c_2)$ will be used to refer to the state at the beginning of the $r^{th}$ iteration. We will also assume for the moment that $c_1, c_2$ are both positive. Using Eq.~\ref{eq:TimeAngleProof1} we have
$ \left| a_1(t_s) \right|^2 = (A c_1)^2 + (B c_2)^2 - 2AB c_1 c_2 \cos(\omega t_s) $ and $ \left| a_2(t_s) \right|^2 = (B c_1)^2 + (A c_2)^2 + 2AB c_1 c_2 \cos(\omega t_s) $. 
Now, note that we can write $2M+1$ as $(2S+1)(A^2 - B^2)$ and therefore it is sufficient to optimize 
\begin{equation}
\resizebox{\FracLineWidth\linewidth}{!}{$
\frac{\braket{S^2}}{(2S+1)} = const + (A^2 - B^2) \left|a_1(t_s)\right|^2 + 2AB\left|a_1(t_s) a_2(t_s)\right|
$}
\end{equation}
where $\braket{S^2}$ is evaluated after the single qubit rotation $R_z(\theta = -\phi)$. Its first derivative with respect to $\omega t_s$ after some algebra can be written as 
\begin{equation} \label{eq:S2slope}
\resizebox{\FracLineWidth\linewidth}{!}{$
\begin{aligned}
\frac{1}{(2S+1)} \frac{\partial \braket{S^2}}{\partial (\omega t_s)} & = 2AB c_1 c_2 \sin(\omega t_s) \\
& \times \left( (A^2 - B^2) - 2AB \frac{\left|a_1(t_s)\right|^2 - \left|a_2(t_s)\right|^2}{\left|a_1(t_s) a_1(t_s)\right|} \right)
\end{aligned}
$}
\end{equation}
Clearly, $t_s = m \pi/\omega$ for some integer $m$, is one of the critical points that nullifies the above derivative. The other critical point can be obtained by setting the term within larger parenthesis to zero. The condition so obtained is $\frac{\left|a_1 (t_s) \right|}{\left|a_2 (t_s) \right|} = \frac{A}{B}$ which is essentially equivalent to $\left|a_1 (t_s) \right| ^2 = A^2$. This corresponds to the desired spin eigenstate. If we write $|a_1(t_s)|^2 = a_1^2[r-1] + 4ABc_1c_2 \sin^2(\omega t_s/2)$, we see that  $|a_1(t_s)|^2$ where $a_1[r-1]=(Ac_1 - Bc_2)$, is incremented in each iteration starting from $a_1[0] = B$. This is the case because $A>B$ for the initial states chosen in the forbidden region $2(S-2M)<1/2$. And hence the above critical point is only reachable in the final iteration. Otherwise, $\left|a_1 (t_s) \right|< A$. 
In other words, the critical point obtained above gives the value of $t_s$ for the last iteration. For all other initial iterations the only least non-zero critical point is $t_s = \pi/\omega$. For these iterations, it turns out this is the only point of maxima and can be verified analysing the sign of slope given by Eq.\ref{eq:S2slope} that depends only on the sign of $\sin(\omega t_s)$ with all other terms being positive (the sign of the term in parenthesis is positive can be established by arguing that $|a_1(t_s)| < A^2$ for all but last iteration).
For this choice of $t_s$, $a_1(t_s)$ and $a_2(t_s)$ end up having opposite sign as long as $Ac_2 - Bc_1>0$ implying $\phi = \pi$. Consequently, the amount of single qubit rotation required is also $\pi$. The condition $Ac_2 - Bc_1 > 0$ is readily satisfied and can be noted from $Ac_2 - Bc_1 > B (>0)$ which in turn can be shown to be equivalent $|a_1|<A$ except for the last iteration. 

Now it remains to justify the rationale for assuming positive $c_1$ and $c_2$. We shall discuss this later but it can be seen that with the choice of $t_s = \pi/\omega$ and $\theta = \pi$, that coefficients at the end of both operation in a given iteration r are real if the coefficients $c_1[r-1]$ and $c_2[r-1]$ are real to start with. It can be established for all iterations since $(c_1(0),c_2(0)) = (B,A)$ are real. This fact forms the basis for the geometric visualization we discuss next where we assume above choices of $t_s$ and $\theta$ unless specified otherwise.

\subsection{Geometric Visualization and map to Grover's Search} 
\label{appsubsec:GroverViz}

Let us consider a (real) plane of points $(c_1,c_2)$ as shown in Fig.~\ref{fig:MapToGrover}. The points on a unit circle can be associated with the states of interest resolved in \{$\ket{X(n,S+1/2,M+1/2}$, $\ket{X(n,S-1/2,M+1/2}$\} basis. The diagonally opposite points on this circle would then correspond to the same state (can be associated with a global phase $\pi$). So, only half the points on this unit circle are sufficient to talk about unique states. We have colored the left-half red and the right-half blue in Fig.~\ref{fig:MapToGrover} to emphasize this fact. 
We will now consider the transformation of the initial state vector $\ket{\psi_i}$ in the first iteration. It is represented by a point$P_0 \equiv (c_1(0),c_2(0))$ as shown in Fig.~\ref{fig:MapToGrover}a. When operated upon by $U(t_s)$, it yields $\ket{\psi(t_s)}$ represented by the point $\overline{P}_0 \equiv (c_1(t_s),c_2(t_s)) = (-c_1(0), c_2(0))$. Therefore the evolution $U(t_s)$ can be seen as a reflection of point $P_1$ about the $C_2$-axis. This can be associated with an oracle $O$ that recognizes the state $\ket{X(n,S+1/2,M+1/2}$ in an arbitrary superposition and marks it by flipping its phase in the superposition. It is usually described by its action on a state $\ket{x}$ as
\begin{equation}
 \ket{x} \xrightarrow[]{O} (-1)^{f(x)} \ket{x} 
\end{equation}
where the function $f(x)$ is defined as 
\begin{multline} 
f(x) = 
\begin{cases}
1 & \text{if} \; \ket{x} = \ket{X(n,S+1/2,M+1/2} \\
0 & \text{otherwise}
\end{cases}
\end{multline}
Note there are no oracle workspace qubits in this realization (can put in summary maybe). Also, note that we are writing the y coordinate first in the ordered pair $(c_1,c_2)$.

The point $\overline{P}_0 \equiv (c_1(t_s),c_2(t_s))$ is equivalently described by the coordinate $(a_1(t_s), a_2(t_s))$ using Eq.~\ref{eq:Ruben3}. Application of $R_z(\theta)$ on $\overline{P}_0$ yields $(-a_1(t_s), a_2(t_s))$ which can be seen as a reflection about the $A_2$ axis. We again write the new coordinate vector back in the original basis using Eq.~\ref{eq:Ruben3} as
\begin{equation} \label{eq:CoordinatesAfterPhaseCorrectionByPi}
\begin{aligned}
c_1(t_2) &= c_1(t_s) - 2 (a_1(t_s)) A \\ 
c_2(t_2) &= c_2(t_s) + 2 (a_1(t_s)) B
\end{aligned}
\end{equation}
where $t_s$ is the moment when we start $R_z$ operation and $t_2$ is the moment when a rotation by $\pi$ is accomplished. The change in coordinate $(c_2(t_2)-c_2(t_s),c_1(t_2)-c_1(t_s))$ is the vector $-2a_1(t_s)(A,-B)$. It should be noted that $(A,-B)$ is a unit vector along $A_1$ axis that is also perpendicular to the initial state $P_0$. Let us write $(A,-B)$ as $\ket{A_1}$, which enables us to write $a_1(t_s) = \braket{A_1|\psi(t_s)}$ and hence the aforementioned change as $-2\ket{A_1}\braket{A_1|\psi(t_s)}$.
The change is thus associated with an operator $-2\ket{A_1}\bra{A_1}$ and the transformed state at time $t_2$, $P_1$, can be obtained by application of the operator $I-2\ket{A_1}\bra{A_1}$ on state at $t_1$ (here $\overline{P}_0$). 
This is clearly the reflection operation used in Grover's search upto a global phase factor of -1. 
\begin{figure}[tbp]
\includegraphics[width=0.48\textwidth, trim={0 0 0 0},clip]{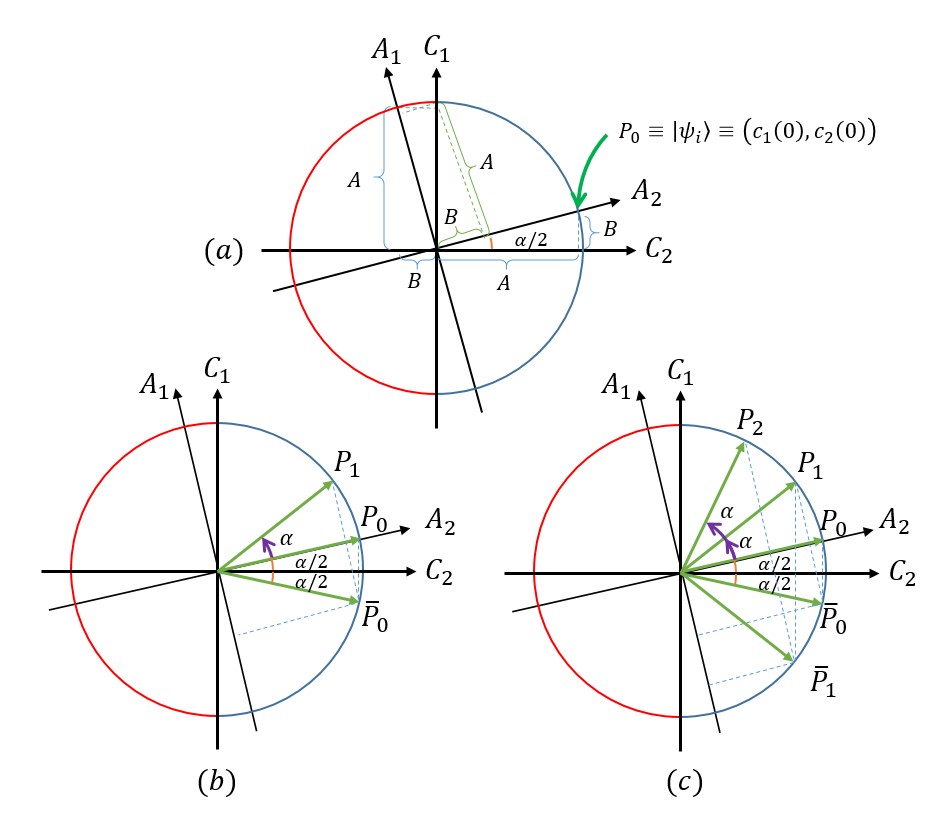} 
\caption [] {\textbf{Visualizating the state transformations}. The state transformations can be visualized as rotations in a 2D plane. The state can be described via coordinates $(c_1,c_2)$ or $(a_1, a_2)$ in either bases as components along $C_1 - C_2$ or $A_1 - A_2$ axes respectively. 
}
\label{fig:MapToGrover}
\end{figure}

The discussion above thus enables us to associate a grover's iterate $(2\ket{A}\bra{A} - I) O$ with the sequence of combined operator in a single iteration, $(I\otimes R_z(\pi)) U(\pi/\omega)$, upto global phases. It should be noted that the transformation of $P_0$ to $P_1$ is a rotation in the $C_1 - C_2$ plane and that the amount of rotation ($\alpha$) in each iteration is double the initial angle the initial state $P_0$ makes with the $C_2$ axis (say $\alpha/2$). The second iteration is quite similar. The point $P_1$ lands up on point $\overline{P}_1$ upon evolution by $U(t_s)$ and then a reflection about $P_0$ yields the point $P_2$. Performing this repeatedly we rotate the vectors $P_{r>0}$ closer and closer to the $C_1$ axis. Since the state vector is rotated discretely (by fixed amount $\alpha$) the final state can land up within a window of angle $\alpha/2$ on either side of $C_1$-axis. If $\ket{\psi_f}$ denotes the final state within this window, the error probability can be associated with $|\braket{X(N,S-1/2,M+1/2)|\psi_f}|^2$. It can be seen that the worst case error probability is $\sin^2(\alpha/2) = \frac{S-M}{2S+1}$. We can chose to stop this rotation before it crosses leaves the first quadrant. And the vector in the final iteration would lie within a window of $\alpha$ angle on the right side of the $C_1$ axis. This decision would thus justify our assumption on the choice of positive $c_1[r]$ and $c_2[r]$ in the previous section but now the worst case error would be $\sin^2(\alpha) = 4\frac{S-M}{2S+1}\frac{S+M+1}{2S+1}$. 
The total number of iterations required can also be determined easily. With the worst case total (in $C_1$-$C_2$ plane) rotation needed (angle subtended by the arc from $P_0$ to the point (1,0)) being $\pi/2$ and the in-plane rotation per iteration being $\alpha = 2 \sin^{-1}(c_1(0))$, one can estimate it to take $O\left(\sqrt{\frac{1}{c_1(0)}}\right) \sim O\left(\sqrt{\frac{2S+1}{S-M}}\right)$ iterations while the precise number of iterations can be written as $\lceil (\pi/2 - \alpha/2)/\alpha \rceil$ where $\lceil \rceil$ denotes the ceiling function.


Until now we discussed evolutions and single qubit rotations by fixed amounts $\pi/\omega$ and $\pi$ that gave us the Grover's iterate, but since these operations are continuous operations, we have a flexibility to choose the interval for evolution and hence the single qubit rotation amount appropriately to reduce the error probability to zero in principle. This choice is already explained in the main text. We will suggest a way to visualize this for which we will first relax considering $c_1$ and $a_1$ to be real. Let's add a third axis orthogonal to the $C_1$-$C_2$ plane of Fig.~\ref{fig:MapToGrover}, where $\operatorname{\mathbb{I}m}\{A_1\}$ or $\operatorname{\mathbb{I}m}\{A_1\}$ can be plotted. So, now the states at any time can be represented by a unit sphere 
We emphasize that this is NOT Bloch sphere but merely an extension of the 2D picture we used before. Now, let's see what the unitary evolution and single qubit rotations mean in this context. For unitary evolution $U(t)$, in $\{\ket{X(n,S+1/2,M+1/2}$, $\ket{X(n,S-1/2,M+1/2} \}$ basis and assuming $c_2$ real, $c_1$ accumulates a factor $e^{-i\omega t}$ over time which can be seen as a clock-wise rotation in $\operatorname{\mathbb{R}e}\{C_1\}$ and $\operatorname{\mathbb{I}m}\{C_1\}$ plane and so the tip of the state vector moves along a circle whose plane is perpendicular to the $c_2$ axis. Similarly, it can be argued that the $R_z(\theta)$ operation, in \{$\ket{X(n-1,S,M} \otimes \ket{0}$, $\ket{X(n-1,S,M+1} \otimes \ket{1}$\} basis, $a_1$ accumulates a factor $e^{-i\theta}$, which implies the state vector, on the sphere described here, moves on a circle that lies on a plane perpendicular to the initial state vector. The initial iterations progresses switching circles whenever they cross the $\operatorname{\mathbb{R}e}\{C_1\}$ - $C_2$ plane. For the final iteration switching of these circles happen when $\operatorname{\mathbb{I}m}\{c_1\} \neq 0$. An example for the case of $\alpha = 0.25\pi/2$ is shown in a supplementary video clip 
at \href{https://youtu.be/0jJju7bsHEU}{https://youtu.be/0jJju7bsHEU}.


\section{Relevant energies of initial state in Generalized Hamiltonian} \label{app:GeneralHamiltonianExplanation}

We claim that the generalized Hamiltonian in k-spin-down subspace described in Eq.~\ref{eq:khotGeneralizedHamiltonian} can be expressed in the following polynomial in $\mathcal{H}_{k,1}$

\begin{equation} \label{eq:polyH}
\mathcal{H}_k = c_0 (\mathcal{H}_{k,1})^0 + c_1 (\mathcal{H}_{k,1})^1 + c_2 (\mathcal{H}_{k,1})^2 + \cdots + c_k (\mathcal{H}_{k,1})^k
\end{equation}

The rationale behind the above equation is as follows: If we apply $H_{k,1}$ on a computational basis state in k-spin-down space, we create states one mismatch in the positions of ones and zeros. If we apply $H_{k,l}$ on a computational basis state in k-spin-down space, we create states $l$ mismatches in the positions of ones and zeros. Terms with $l$ mismatches can also be obtained by applying $H_{1,k}$ $l$ times. It is therefore possible to write $H_{k,l}$ in terms of polynomial powers of $H_{k,1}$.

We will get the expressions for the $c_l$ coefficients  in the following but with this expression we can readily obtain the eigenvalues of $\mathcal{H}_k$ in terms of the eigenvalues of $\mathcal{H}_{k,1}$. To simplify the algebra, we define matrices $M_{k,l}$ as, $\mathcal{H}_{k,l} = 2J_l M_{k,l}$.  Thus the matrices $M_{k,l}$ consists of only 0's and 1's. The generalized Hamiltonian in k-spin-down subspace described in Eq.~\ref{eq:khotGeneralizedHamiltonian} can be written as  $\mathcal{H}_k = \sum_{l = 0}^k 2J_l M_{k,l}$.




First, we will show by induction that $(M_{k,1})^l$ can be expanded as ($l \leq k$)

\begin{equation} \label{eq:InductionStep1}
(M_{k,1})^l = a_0^{[l]} M_{k,0} + a_1^{[l]} M_{k,1} + a_2^{[l]} M_{k,2} + \cdots a_l^{[l]} M_{k,l}
\end{equation}
 
with $(M_{k,1})^0 := M_{k,0}$. Clearly, $a_0^{[0]} = 1$. Also, $(M_{k,1})^1 = 0*M_{k,0} + 1*M_{k,1}$ i.e. $a_0^{[1]} = 0$ and $a_1^{[1]} = 1$. Here, the superscripts in square brackets are used for notational convenience and should not be confused with powers. Now, left multiplication by $M_{k,1}$ on $(M_{k,1})^{l-1}$ yields

\begin{equation} \label{eq:InductionStep2}
\begin{aligned}
M_{k,1} (M_{k,1})^{l-1}= & a_0^{[l-1]} M_{k,1} M_{k,0} + a_1^{[l-1]} M_{k,1} M_{k,1} \\ + & a_2^{[l-1]} M_{k,1} M_{k,2} + \cdots a_{l-1}^{[l-1]} M_{k,1} M_{k,l-1}
\end{aligned}
\end{equation}

Above equation contains terms like $M_{k,1} M_{k,q}$. For simplification of these terms, consider a state $\psi_1 = \ket{11...1,00...0}$ where ',' separates k ones on the left from n-k zeros on the right. Application of $M_{k,q}$ on $\psi_1$ produces a state $\psi_2$ which is a superposition of computational basis states $\ket{u_i}$ in k-spins-down subspace with literals $u_i$ differing at q positions from that of $\psi_1$ ($u_i$ has k-q ones in the left chunk before comma while q ones on the right chunk after comma). Likewise application of $M_{k,1}$ on $\psi_2$ yields a state  $\psi_3$ composed of $\ket{u_j}$ with one mismatch in the literals $u_j$ and $u_i$. It should be noted therefore, $\psi_3$ consists of states differing at $q-1$, $q$ and $q+1$ positions from $\psi_1$ which can equivalently be produced from application of $M_{k,q-1},M_{k,q}$ and $M_{k,q+1}$ on $\psi_1$. One can thus decompose $M_{k,1} M_{k,q}$ as

\begin{equation} \label{eq:InductionStep3}
M_{k,1} M_{k,q} = \alpha_q M_{k,q-1} + \beta_q M_{k,q} + \gamma_q M_{k,q+1}
\end{equation}

Now since there are a total of ${}^kC_q \cdot {}^{n-k}C_q$ terms in $\psi_2$ and each $\ket{u_i}$ comprising $\psi_2$ can produce $^qC_1 \cdot ^qC_1$ (shift a 1 from the right chunk to the left) terms that mismatch with $\psi_1$ at $q-1$ positions, a total of ${^qC_1 \cdot ^qC_1 \cdot {}^kC_q \cdot {}^{n-k}C_q}$ terms are formed (with overcounting). It can be checked that this is larger than the total number of terms that can be produced on application of $M_{k,q-1}$ on $\psi_1$. The over-counted terms are actually distributed uniformly among the $^kC_{q-1} \cdot ^{n-k}C_{q-1}$ terms in $M_{k,q-1} \psi_1$ providing the value of $\alpha_q = (k-q+1)*(n-k-q+1)$. One can similarly obtain $\beta_q=q(n-2q)$ and $\gamma_q = (q+1)^2$. Therefore, using Eq.~\ref{eq:InductionStep2} and Eq.~\ref{eq:InductionStep3}, we can obtain Eq.~\ref{eq:InductionStep1} with the following expression of coefficients 

\begin{equation} \label{eq:a_s_relation}
a_q^{[l]} = a_{q-1}^{[l-1]} \gamma_{q-1} + a_{q}^{[l-1]} \beta_q + a_{q+1}^{[l-1]} \alpha_{q+1}
\end{equation}

where $ q \leq l \leq k$. This relation provides us with an iterative procedure to obtain the 'a' coefficients on the RHS of equation (\ref{eq:InductionStep1}) of $M_l$ in the expansion of $M_1^l$. Let us also define $a_{q>l}^{[l]} = 0$. This enables us to arrange Eq.~\ref{eq:a_s_relation} in the following form

\begin{equation} \label{eq:Obtain_a_s}
\begin{bmatrix} 
\beta_0  & \alpha_1 & 0        & 0        & \cdots & 0            & 0            & 0 \\ 
\gamma_0 & \beta_1  & \alpha_2 & 0        & \cdots & 0            & 0            & 0 \\ 
0        & \gamma_1 & \beta_2  & \alpha_3 & \cdots & 0            & 0            & 0 \\ 
0        & 0        & \gamma_2 & \beta_3  & \cdots & 0            & 0            & 0 \\ 
\vdots   & \vdots   & \vdots   & \vdots   & \ddots & \vdots       & \vdots       &  \vdots\\ 
0        & 0   	    & 0 & 0   	    &        & \beta_{k-2}  & \alpha_{k-1} & 0 \\ 
0 		 & 0   	    & 0 & 0   	    & \cdots & \gamma_{k-2} & \beta_{k-1}  & \alpha_{k} \\ 
0 		 & 0   	    & 0 & 0   	    & \cdots & 0            & \gamma_{k-1} & \beta_{k} \\ 
\end{bmatrix} 
\begin{bmatrix} 
a_0^{[l-1]} \\ 
a_1^{[l-1]} \\ 
a_2^{[l-1]} \\ 
     \\ 
\vdots \\ 
    \\ 
a_{k-1}^{[l-1]} \\ 
a_{k}^{[l-1]} \\ 
\end{bmatrix}
=
\begin{bmatrix} 
a_0^{[l]} \\ 
a_1^{[l]} \\ 
a_2^{[l]} \\ 
     \\ 
\vdots \\ 
    \\ 
a_{k-1}^{[l]} \\ 
a_{k}^{[l]} \\ 
\end{bmatrix}
\end{equation}

Let's call the $(k+1) \times (k+1)$ matrix on the left as $A$. It can be seen that for a given l, $a^{[l-1]}_{l-1< q \leq k} = 0$ and only after application of $A$, yields a non-zero $a^{[l]}_{q=1} = 0$ and $a^{[l]}_{q>l}$ still remains zeros respecting the definition. It should also be noted that $A$ is constant for a given n and k. So, starting from a column vector containing $a_0^{[0]} = 1$ and $a_{q>0}^{[0]}$, one can directly obtain the coefficients $a_q^{[l]}$ on left multiplication by $A^l$. 

This completes the proof for Eq.~\ref{eq:InductionStep1}. Now, it can be readily seen that, since each $M_{k,1}^l$ is a linear combination of $M_{k,i<l}$, Eq.~\ref{eq:polyH} entails that $\mathcal{H}_{k,l}$ is also a linear combination of $M_{k,i<k}$. We can collect the appropriate terms with Eq.~\ref{eq:InductionStep1} substituted in Eq.~\ref{eq:polyH} and compare with coefficients of $M_{k,l}$ in Eq.~\ref{eq:khotGeneralizedHamiltonian} (substituted with $\mathcal{H}_{k,l} = 2J_l M_{k,l}$) to obtain the following

\begin{equation} \label{eq:Obtain_c_s}
\begin{bmatrix} 
a_0^{[0]} & a_0^{[1]} & a_0^{[2]} & \cdots       & a_0^{[k-1]}      & a_0^{[k]} \\ 
0 & a_1^{[1]} & a_1^{[2]} & \cdots       & a_1^{[k-1]}      & a_1^{[k]} \\ 
0 & 0         & a_2^{[2]} & \cdots       & a_2^{[k-1]}      & a_2^{[k]} \\ 
\vdots  & \vdots    & \vdots    & \ddots       & \vdots           & \vdots    \\ 
0 & 0   	  & 0   	  & \cdots       & a_{k-1}^{[k-1]}  & a_{k-1}^{[k]} \\ 
0 & 0   	  & 0   	  & \cdots       & 0       & a_k^{[k]} \\ 
\end{bmatrix} 
\begin{bmatrix} 
c_0 \\ 
c_1 (2J_1)^1 \\ 
c_2 (2J_1)^2 \\ 
\\
\vdots \\ 
\\
c_{k-1} (2J_1)^{k-1} \\ 
c_k (2J_1)^k \\ 
\end{bmatrix}
=
\begin{bmatrix} 
2J_0 \\ 
2J_1 \\ 
2J_2 \\ 
     \\ 
\vdots \\ 
    \\ 
2J_{k-1} \\ 
2J_{k} \\ 
\end{bmatrix}
\end{equation}

It can be seen that the columns of the $(k+1) \times (k+1)$ matrix above can be obtained from Eq.~\ref{eq:Obtain_a_s}. One can thus solve for $c_i's$ and obtain the relevant energies Eq.~\ref{eq:polyH}. If $E_1$($E_2$) is an eigenvalue $\mathcal{H}_{k,1}$ then the eigenvalue of $\mathcal{H}_{k}$ is simply the same polynomial in Eq.~\ref{eq:polyH} with $\mathcal{H}_{k,1}$ replaced by $E_1$($E_2$).

Eq.~\ref{eq:Obtain_a_s} and Eq.~\ref{eq:Obtain_c_s} are useful for numerical evaluations but we can easily obtain the required eigenvalues for smaller values of $k$. 
Using Eq.~\ref{eq:InductionStep1}, we can write $M_{k,2}=[(M_{k,1})^2-(n-2)(M_{k,1})^1-k(n-k)(M_{k,1})^0]/4$. Using this we can obtain eigenvalues of $M_{k,2}$ from eigenvalues of $M_{k,1}$. Further eigenvalues of $M_{k,1}$ are given by: $S(S+1)-(3/4)n-(1/2)[ ^k C_2 + ^{n-k} C_2 - k(n-k)]$, where the last factor is square bracket is the diagonal term stated below \ref{eq:khotHamiltonian}. As discussed before $S$ takes values from $n/2$ to $n/2-k$, giving  $k+1$ distinct eigenvalues. Recalling the relation $\mathcal{H}_{k,l} = 2J_l M_{k,l}$, the eigenvalues of $H_{k,2}$ are simply $2J_2$ times eigenvalues of $M_{k,2}$. In the two spin-down subspace (i.e. $k=2$), the eigenvalues of $H_{2,1}$ are $4(n-2)J_1, 2(n-4) J_1$ and $-4J_1$. The corresponding eigenvalues of $H_{2,2}$ are $(n-2)(n-3)J_2, 2(3-n)J_2$ and $2J_2$. The difference between the first two eigenvalues (since only these are correspond to the basis states involved in evolution), $n(n-3)J_2$ determines the time required for entangling evolution using $H_{2,2}$, while preparing Dicke states (with 2 spins in down state).

\section{More on cost of Preparation of Dicke states using modified Algorithm} \label{app:Sublinearity}
Based on discussion in Appendix.~\ref{appsubsec:GroverViz}, we can write the exact cost of preparing $D^p_q$ from $D^{p-1}_{q-1}$, $r_{(p,q)}$ as $\lceil \frac{\pi}{4} \sqrt{\frac{p}{q}} -\frac{1}{2} \rceil$ which we shall write as $\frac{\pi}{4} \sqrt{\frac{p}{q}} +\frac{1}{2}$ to estimate the order of total cost of preparing $D^n_k$ expanding a W-state. We choose to write $n-k = n_0$, which enables us to write $\sum_{j=2}^{k} r_{(n-k+j,j)}$ from Eq.~\ref{eq:TotalCost} as  $\sum_{q=2}^{k} \left( \frac{\pi}{4} \sqrt{1+\frac{n_0}{q}} +\frac{1}{2} \right)$ and let us call this as $S_k$. Note that the requirement $k\leq n$ translates to $k\leq n_0$ now. Using the fact that the arithmetic mean of positive real numbers is less than or equal to the quadratic mean, we can write:
\begin{equation}
S_k \leq \frac{\pi}{4} \sqrt{k-1} \left( n_0 \sum_{q=2}^{k} 1/q + (k-1) \right)^{1/2} + \frac{k-1}{2}
\end{equation}
Now utilizing the fact that $\sum_{q=2}^{k} 1/q < \int_1^k \frac{1}{x} dx = \log{k}$ and ignoring the $-1$'s with $k$'s in the above equation we have
\begin{equation}
S_k < \frac{\pi}{4} k \left( 1 + \frac{n_0}{k} \log{k} \right)^{1/2} + \frac{k}{2}
\end{equation}
We can estimate the behaviour of right-hand-side (RHS) of above equation for $k=n_0^a$ in the limit of large $n$.
Note $k>1$ implies $a > 0$ while $k\leq n_0$ implies $a\leq 1$ but we shall restrict the discussion to $0<a<1$ since no significant conclusion is obtained for $a=1$ using the inequality above. To show that RHS is better than $O(n)$ it is sufficient to show that the first term of RHS is better than $O(n_0)$ since $n=n_0+k$. As such in the limit of large $n_0$ we have the cost equal to $O\left( n_0^{(1+a)} \log{n_0} \right)^{1/2}$ which can be argued to be better than $O(n_0)$. We know that the logarithm function grows slowly than any power function and choosing a power function as $n_0^{(1-a)}$ we can easily establish the previous assertion since $\log(n_0) < n_0^{(1-a)}$ for sufficiently large $n_0$.
Although we have written the result in terms of $n_0$ and $a$ defined previously, it can be seen that this cost is sub-linear with n for the select choice of $k$ noted above.

A more precise calculation for the cost for general k can be made, that we shall only suggest here. Since $r_{(p,q)} = 1$ in the pink region, the cost accumulated in the summation reduces to 1. It can be argued that this happens when $q = \lfloor n_0/3 \rfloor (= q' say)$ and hence $S_k = S_{q'} + (k-q')$. Clearly, the increment in the cost beyond $q'$ becomes linear with $k$. Since, the total cost of linear step algorithm is $n_0 + k = (n_0+q') + (k-q')$, the cost of the modified algorithm is better if $S_{q'} < n_0 + q'$. This turns out to be true as seen in Fig.~\ref{fig:TimeComplexity} and can also be shown by noting $S_{q'}< \int_1^{q'}{(\frac{\pi}{4}\sqrt{1+\frac{n_0}{x}} + \frac{1}{2})dx}$. The integral can be obtained in closed form and in the limit of large $n_0$ can be approximated as $\frac{\pi}{4} \left(\frac{2}{3} + \frac{1}{2} \log{(3)} \right) n_0 + \frac{1}{2}(\frac{n_0}{3}-1)$, where we have put $q' = n_0/3$ i.e have also included its fractional part. Specific values substituted it turns out to be less than $4n_0/3$ completing the proof and justifying Fig.~\ref{fig:TimeComplexity}.

\end{document}